\documentclass[aps,preprint,nofootinbib,eqsecnum]{revtex4}%
\usepackage{amsmath}
\usepackage{amsfonts}
\usepackage{amssymb}
\usepackage{graphicx}
\usepackage{amscd}%
\allowdisplaybreaks
\begin{document}

\begin{flushleft}
USCHEP/0311ib6
\hfill hep-th/0311264 \\
CERN-CH/2003-285 \\
OHSTPY-HEP-T-03-014 \vskip 1cm
\end{flushleft}

\title{Improved Off-Shell Scattering Amplitudes \\in String Field Theory\\and New Computational Methods \bigskip}
\author{Itzhak Bars}
\affiliation{Department of Physics, USC, Los Angeles, CA
90089-0484\\ Theory Division, CERN, CH-1211 Geneva 23,
Switzerland}
\author{I.Y. Park}
\affiliation{Department of Physics, The Ohio State University, Columbus, OH 43210, USA \vskip 0.75cm}

\begin{abstract}
We report on new results in Witten's cubic string field theory for the
off-shell factor in the 4-tachyon amplitude that was not fully obtained
explicitly before. This is achieved by completing the derivation of the
Veneziano formula in the Moyal star formulation of Witten's string field
theory (MSFT). We also demonstrate detailed agreement of MSFT with a number of
on-shell and off-shell computations in other approaches to Witten's string
field theory. We extend the techniques of computation in MSFT, and show that
the j=0 representation of SL(2,R) generated by the Virasoro operators
$L_{0},L_{\pm1}$ is a key structure in practical computations for generating
numbers. We provide more insight into the Moyal structure that simplifies
string field theory, and develop techniques that could be applied more
generally, including nonperturbative processes.

\end{abstract}
\maketitle
\tableofcontents

\newpage

\section{Introduction}

In this paper we will report on new results in Witten's string field theory
\cite{witten} for the off-shell 4-tachyon amplitude that was not fully
computed explicitly before. In previous computations the off-shell factor
$f\left(  x\right)  $ was obtained in an implicit form \cite{samuel}%
\cite{sloan}, while an explicit computation reported only the first two terms
in an expansion in the vicinity of the integration limit of the parameter~$x$
\cite{sloan}. These results were obtained by using mainly conformal mapping
techniques \cite{mandelstam} and followed the methods of Giddings' original
computation \cite{giddings} of the on-shell Veneziano amplitude, which could
be extended to off-shell under the guidance of the oscillator formulation of
string field theory \cite{CST}\cite{GJ}\cite{ohta}. In this paper we will
obtain a fairly comprehensive profile of the off-shell factor $f\left(
x\right)  $ in its entire range $0\leq x\leq1$ by giving the explicit form to
tenth order in the parameter $x$ (first two terms in agreement with
\cite{sloan}), obtaining a plot in the full range, and computing the critical
slope at a turning point in midrange that determines differentiability.
Contrary to our result, previously it was thought that the function $f\left(
x\right)  $ was not differentiable at $x=1/2,$ where it was not well
understood. These are achieved in the Moyal star formulation of Witten's
string field theory (MSFT) \cite{B1}-\cite{BKM3}.

The usual approach of computation starts with a precise formulation of
Witten's string field theory, such as the oscillator formulation, to derive a
formal expression for a string Feynman diagram in terms of the cubic vertex
defined in terms of the Neumann coefficients. After this step a jump is made
to conformal maps from an analog model \cite{mandelstam} and the real
computation is performed by using conformal field theory, if the conformal map
can be constructed. The desired conformal maps can be found explicitly only in
certain lucky cases, and the four point function is one of them. The conformal
map procedure has been used virtually in every successful analytic
computation, while the oscillator basis is directly pursued mainly with
numerical studies using level truncation \cite{KP}\cite{taylor} because of the
complexity of the Neumann coefficients. These have been some of the
challenging features of Witten's string field theory in various formulations
\cite{GJ}\cite{peskin} that, despite the beauty of Witten's basic action, have
led to limited results in string field theory.

Some of the complexities of other approaches are not present, or take an
easier form in MSFT. In this formulation the string joining star product is
the simple Moyal product, and this reproduces directly results in string field
theory, in agreement with conformal field theory, but without mapping back to
conformal field theory or other intermediate steps. Therefore some of the
lingering problems in string field theory seem to be good testing grounds for MSFT.

In this paper we apply the MSFT techniques to the off-shell 3-point and
4-point amplitudes. We will derive the Veneziano amplitude directly from MSFT
and determine the off-shell factor far more accurately compared to previous
computations. In this process we demonstrate that MSFT agrees in detail with
other approaches while bringing efficiency into the computations, and also
obtain new results in MSFT that other approaches could test. We view these
computations as a preparation for our ultimate aim which is the investigation
of nonperturbative string theory phenomena by using the simpler MSFT. We
believe the techniques and insights developed in this paper will be useful for
this purpose. In fact we find that some of the new results and insight gained
by this work impact the way to compute nonpertubative quantities in string
field theory as discussed in section VI.

In the rest of this section we will describe briefly MSFT and introduce some
notation. In section II we discuss off-shell 3-point functions. This is
necessary to understand the behavior of the theory with respect to a cutoff in
mode space, and to define the physical string coupling as opposed to the bare
divergent coupling that appears in the action. The results we obtain for the
off-shell 3-point amplitudes demonstrate detailed agreement between MSFT and
conformal field theory or the oscillator formulation of string field theory.

In section III we analyze the off-shell 4-tachyon amplitude. We show directly
from MSFT, without connecting through conformal field theory, that on-shell we
obtain the Veneziano amplitude. Furthermore we obtain the off-shell factor
$f\left(  x\right)  $ and compute the first ten terms in an expansion in
powers of $x$, provide a plot in its full range $0\leq x\leq1$, and determine
its differentiability at a turning point. Previously it was thought that the
derivative was discontinuous at mid-range and only the lowest two terms were
explicitly computed \cite{sloan} (in agreement with our result). Our
computation provides the most comprehensive information on the off-shell
4-tachyon amplitude produced so far in string field theory.

In section IV we develop the details of the tools that allowed us to perform
the computations in sections II and III and which would be applicable more
generally to other computations.

In section V we discuss the SL$\left(  2,R\right)  $ generated by the Virasoro
generators $L_{0},L_{\pm1},$ whose $j\left(  j+1\right)  =0$ representation
deeply underlies the structures that appear in our computations. By using some
group theoretic properties of this very special representation of SL$(2,R)$ we
develop tools for computation in string field theory that are needed in our
paper to generate numbers. We also compare the discrete and continuous Moyal
bases which are simply two different bases of the $j=0$ representation.

In section VI we discuss some of the impact that our present computations have
on the non-perturbative landscape, and then conclude in section VII. In the
Appendix we give further results produced through the techniques in section
IV, and relations to Neumann coefficients.

The action of Witten's cubic string field theory \cite{witten} in the MSFT
formalism in the Siegel gauge is
\begin{equation}
S\left(  A\right)  =-\int d^{d}\bar{x}\,Tr\left(  \frac{1}{2\alpha^{\prime}%
}A\star(L_{0}-1)A+\frac{g_{0}}{3}A\star A\star A\right)  .\label{action}%
\end{equation}
where $L_{0}$ is given in Eqs.(\ref{L0},\ref{L0gh}). The zero mode ghosts have
already been dealt with \cite{BKM3} so they no longer appear. Hence the field
$A$ here is equivalent to the physical field in Witten's theory. The string
field $A\left(  \bar{x},x_{e},p_{e}\right)  $ is written in a mixed
position-momentum basis, which is equivalent to a Fourier transform of the
purely position basis. In this basis string joining is represented by the
usual Moyal star product \cite{B1} in a noncommutative phase space $\left(
x_{e},p_{e}\right)  $ labelled by the even modes $e=2,4,6,\cdots$
\footnote{The position basis is given in terms of the even and odd modes
$\left(  x_{e},x_{o}\right)  .$ The Fourier transform in the odd modes maps to
the space $\left(  x_{e},p_{o}\right)  .$ Witten's string joining star product
becomes a non-diagonal Moyal product in the space $\left(  x_{e},p_{o}\right)
$ \cite{B1}. This is diagonalized by defining $p_{e}$ as an infinite
combination of the $p_{o}=p_{e}T_{eo}$ by introducing the special matrix
$T_{eo}$ given in Eq.(\ref{TR}). In this way we arrive to the noncommutative
space $\left(  x_{e},p_{e}\right)  $ with a diagonal Moyal product whose
meaning is string joining. We emphasize that this $p_{e}$ is a definition in
terms of $p_{o},$ it should not be confused with the first quantized momentum
that is canonical to $x_{e},$ whose representation in this space is the
derivative $i\partial_{x_{e}}.$ However, interestingly this can be reproduced
in the noncommuative geometry (string joining) relation $-i\partial_{x_{e}%
}A\left(  \bar{x},x_{e},p_{e}\right)  =\left[  p_{e},A\left(  \bar{x}%
,x_{e},p_{e}\right)  \right]  _{\star}.$ A closely related continuous Moyal
basis \cite{DLMZ} is obtained by orthogonality transformations, as will be
discussed later in the paper.}. The star product $\star$ is local in the
midpoint variable $\bar{x},$ and is independent for each $e.$ This separation
of variables and the simplicity of the Moyal star are the main conceptual and
practical simplifications that lead to new progress by overcoming midpoint
problems in other approaches and opening up easier computational techniques in MSFT.

It has been shown in \cite{B1}-\cite{BKM3} through some explicit computations
that MSFT is in full agreement with other computational approaches to Witten's
string field theory, including the oscillator formulation and conformal field
theory \cite{GJ}-\cite{peskin}. In particular, the Moyal star reproduces the
Neumann coefficients that define the vertices in the oscillator formulation of
string field theory \cite{BM2}\cite{BKM3}. Furthermore, the MSFT propagator
has the usual free string spectrum. Therefore, even though it is a very
different computational formalism, due to the one to one correspondence
described in \cite{BKM1} we expect identical final results between MSFT and
oscillator approach computations of any string Feynman graph. This expectation
will be confirmed in detail in this paper. This demonstrates once again, and
in greater detail, that MSFT is a precise representation of Witten's string
field theory.

The string in 26 dimensions is supplemented with two additional fermionic
dimensions that describe the conformal ghosts $b,c,$ with the appropriate
generalization of the Moyal star product for fermions. The traditional
perturbative string states (tachyon, vector, etc.) are identified through the
usual expansion%
\begin{equation}
A\left(  x_{cm},x_{e},p_{e}\right)  =T(x_{cm})A_{0}\left(  x_{e},p_{e}\right)
+V_{\mu}(x_{cm})\left(  \alpha_{-1}^{\mu}A_{0}\left(  x_{e},p_{e}\right)
\right)  +....
\end{equation}
where $A_{0}\left(  x_{e},p_{e}\right)  $ is the perturbative vacuum string
field configuration and $\alpha_{-n}^{\mu}$ is a differential operator
representation of string oscillators in the space $\left(  x_{e},p_{e}\right)
$ \cite{BM2}\cite{BKM3}. $A_{0}$ is a specific normalized Gaussian $Tr\left(
A_{0}\star A_{0}\right)  =1$, that represents the vacuum $L_{0}A_{0}=0,$
including ghosts \cite{BKM3} (see also \cite{erler}). It is given by%
\begin{equation}
A_{0}\left(  \xi\right)  =\left\vert \det4m_{0}\right\vert ^{d/4}\left\vert
\det4m_{0}^{-1}\right\vert ^{-2/4}~\exp\left(  -\eta_{\mu\nu}\xi^{\mu}%
m_{0}\sigma^{-1}\xi^{\nu}+i\varepsilon_{mn}\xi^{m}m_{0}^{-1}\sigma^{-1}\xi
^{n}\right)  ,\;\;d=26,
\end{equation}
where $\xi^{\mu},\xi^{m}$ are the non-commutative coordinates written as a
doublet for each $e$, with the bosonic part $\xi_{i}^{\mu}=\left(  x_{e}^{\mu
},p_{e}^{\mu}\right)  $ for matter and fermionic part $\xi^{1}=(x_{e}%
^{b}/\sqrt{2\alpha^{\prime}},~-\sqrt{2\alpha^{\prime}}p_{e}^{c}),~\xi
^{2}=(x_{e}^{c}/\sqrt{2\alpha^{\prime}},~\sqrt{2\alpha^{\prime}}p_{e}^{b})$
for the $b,c$ ghosts\footnote{Relative to \cite{BKM2}\cite{BKM3} we are
improving notation by introducing $\xi^{m},$ with $m=1,2,$ and the Sp$\left(
2\right)  $ metric $\varepsilon_{mn},$ for the sake of connecting with an
upcoming paper in which we will discuss some useful hidden symmetries that
connect matter and ghosts in the MSFT formalism, and further simplify the
structure and computations.}. Each pair $\left(  x,p\right)  $ satisfies
standard commutation/anticommutation rules under the star product. These can
be written compactly as $[\xi_{i}^{\mu},\xi_{j}^{\nu}]_{\star}=\eta^{\mu\nu
}\sigma_{ij}$ for matter and $\left\{  \xi_{i}^{m},\xi_{j}^{n}\right\}
_{\star}=-i\varepsilon^{mn}\sigma_{ij}$ for ghosts, where the symbol
$\sigma=-\theta\sigma_{2}$ is the Pauli matrix $\sigma_{2}$ in the doublet
space multiplied with a noncommutativity parameter $\theta.$ We take
$\theta=1$ by a choice of units. The $m_{0}$ that appears in the vacuum state
$A_{0}\left(  \xi\right)  $ is a matrix in mode space determined by
$L_{0}A_{0}=0$ as shown in \cite{BM2}. Although $m_{0}$ is a simple matrix, we
will not need it explicitly in this paper.

The perturbative particle fields $T(x_{cm}),V_{\mu}(x_{cm}),$ etc. are
expressed as functions of the center of mass $x_{cm}.$ The star product
$\star$ is local in the midpoint of the string $\bar{x}$, not in the center of
mass $x_{cm}$. Therefore, before evaluating the interaction for any
perturbative field $T(x_{cm}),V_{\mu}(x_{cm}),$ etc. one must first write the
center of mass in terms of the midpoint $x_{cm}=\bar{x}+w_{e}x_{e}$, where
$w_{e}=-\sqrt{2}\left(  -1\right)  ^{e/2}$. This is a crucial step in
computations of the star products. The midpoint had been a source of numerous
problems in the split string formalism, and the resolution first was given in
the context of MSFT in \cite{BM2} with the simple prescription just described.

In the quadratic term the star product plays no role, and could be removed, as
is usual for the Moyal product. The coefficient of the quadratic term in the
action is chosen such that the particle fields are correctly normalized (after
taking into account the definitions of the trace $Tr$ including ghosts and the
Virasoro operator $L_{0}$ as given in \cite{BKM3})
\begin{align}
S_{quadratic}  &  =-\int d^{d}\bar{x}\left(  \frac{1}{2}\partial_{\mu
}T\partial^{\mu}T+\frac{1}{4}F_{\mu\nu}F^{\mu\nu}+\cdots\right) \\
&  =\int d^{d}\bar{x}\frac{1}{2}\left(  \partial_{0}T\partial_{0}T-\vec
{\nabla}T\cdot\vec{\nabla}T\right)  +\frac{1}{2}\partial_{0}V^{\mu}%
\partial_{0}V_{\mu}+\cdots
\end{align}
Then from the cubic term one finds the Feynman rules and computes the
amplitudes as described in \cite{BKM1}\cite{BKM3}. For each external line we
insert a string field representative of a particle. In particular an incoming
tachyon with momentum $k^{\mu}$ is represented by the string field
$A_{0}\left(  x_{e},p_{e}\right)  e^{ik\cdot x_{cm}},$ where $A_{0}$ is the
normalized vacuum field given above, and $e^{ik\cdot x_{cm}}$ is the center of
mass plane wave which is part of $T\left(  x_{cm}\right)  $. To compute its
interactions one must write it in the form $A_{0}\left(  x_{e},p_{e}\right)
e^{ik\cdot\left(  \bar{x}+w_{e}x_{e}\right)  }$ which is a shifted Gaussian in
$\left(  x_{e},p_{e}\right)  $ space. These details are fully explained in
\cite{B1}-\cite{BKM3}, where it is also shown how to compute the star product
and the trace with efficient methods based on a monoid algebra of shifted
Gaussians \cite{BM2}.

In the expressions below the constant matrix $t_{eo}$ and vectors $w_{e}%
,v_{o}$ in even/odd mode space (and related matrices $T_{eo},$ $R_{oe}$) are
fundamental matrices in MSFT that encode the joining of strings \cite{B1}%
-\cite{DLMZ}, and are given explicitly in Eqs.(\ref{v},\ref{TR}). The matrices
$\kappa_{o},\kappa_{e}$ are diagonal matrices that represent the odd/even
oscillator frequencies $\kappa_{o}=diag\left(  o\right)  ,$ $o=1,3,5,\cdots$
and $\kappa_{e}=diag\left(  e\right)  ,$ $e=2,4,6,\cdots$ as in
Eq.(\ref{kappa}) below. A bar on top of a square or column matrix symbol, such
as $\bar{t},\bar{v},$ etc. means the transpose of the matrix. In certain
computations, to avoid associativity anomalies these infinite matrices must be
replaced with their regulated $N\times N$ version as given in \cite{BM1}%
\cite{BKM2}\cite{BKM1}\cite{BKM3} and then $N$ must be sent to infinity at the
end. The regulated matrices obey some nice algebraic properties which are also
shared by the infinite matrices, thereby permitting analytic computation in
the finite $N$ version. A particular form of the regulator which we have found
useful in some computations is given in footnote \ref{regulator}.

It is necessary to use the regulator in those computations where we suspect
anomalous behavior, but otherwise the unregulated matrices can be used, as we
will do for most of the computations in this paper. It is interesting to note
that the regulated matrices with only a few modes (small $N$) reproduce
approximately most of the numerical results we obtain with more sophisticated
methods at $N=\infty$. An example is the Neumann matrices as given in the
appendix of \cite{BKM3}, and many of the numbers computed in the current
paper, although we do not make the effort to demonstrate this point in this paper.

\section{Off-Shell 3-point Functions}

In this section first we briefly outline the off-shell 3-point functions to
establish the relation between the bare coupling $g_{0}$ that appears in the
action of MSFT and the on-shell tachyon coupling $g,$ which is identified with
the string coupling. The relation between the two involves a factor which
diverges with the number of modes $2N$ as $g_{0}\sim\left(  2N\right)
^{3/2}g$ as will be explained later. All other scattering amplitudes are
proportional to some power of the bare coupling $g_{0},$ and this must be
first written in terms of the finite on-shell tachyon coupling $g$. After this
step, it is seen that all amplitudes are finite. In this process one finds a
renormalization factor in front of the amplitudes multiplying a power of the
on-shell coupling $g$. An example of this is the factor $g^{2}\left(
27/16\right)  ^{3}/4$ for the 4-tachyon amplitude in Eq.(\ref{a4}) below.

The off-shell 3-tachyon amplitude is obtained by using the Feynman rules (note
$\frac{1}{3}g_{0}\times3!=2g_{0}$) and inserting the tachyon field for the
external leg in the Feynman graph, leading to the expression%
\begin{equation}
g_{123}\left(  k_{i}\right)  =2g_{0}\int d^{d}\bar{x}Tr\left(  A_{0}%
e^{ik_{1}\cdot\left(  \bar{x}+wx_{e}\right)  }\star A_{0}e^{ik_{2}\cdot\left(
\bar{x}+wx_{e}\right)  }\star A_{0}e^{ik_{3}\cdot\left(  \bar{x}%
+wx_{e}\right)  }\right)  . \label{g123int}%
\end{equation}
The $\bar{x}$ integral gives a momentum conservation delta function $\left(
2\pi\right)  ^{26}\delta^{\left(  26\right)  }\left(  k_{1}+k_{2}%
+k_{3}\right)  $, while the trace and star products are easily evaluated by
using the bose/fermi monoid rules developed in \cite{BM2}\cite{BKM3}. The
result is
\begin{equation}
g_{123}\left(  k_{i}\right)  =g~\left(  \frac{27}{16}\right)  ^{\frac{3}%
{2}-\frac{1}{2}\alpha^{\prime}\left(  k_{1}^{2}+k_{2}^{2}+k_{3}^{2}\right)  },
\label{g123}%
\end{equation}
multiplied with the delta function. This is in full agreement with previous
studies of the off-shell 3-tachyon amplitude \cite{CST}\cite{GJ}\cite{samuel}.
It gives precisely $g$ on shell by definition $g_{123}\left(  k_{i}\right)
|_{\alpha^{\prime}k_{i}^{2}=1}=g.$ The remaining off-shell factor in our
calculation initially has the form $\exp\left(  \omega\left(  \frac{3}%
{2}-\frac{1}{2}\alpha^{\prime}\left(  k_{1}^{2}+k_{2}^{2}+k_{3}^{2}\right)
\right)  \right)  ,$ with $\omega=\frac{1}{2\alpha^{\prime}}\left(
w,0\right)  \sigma m_{0}\left(  3+m_{0}^{2}\right)  ^{-1}\left(  w,0\right)
^{T}. $ This is obtained easily by evaluating the star products and trace in
Eq.(\ref{g123int}) by using the simple monoid methods in \cite{BM2}%
\cite{BKM3}. Using the definition of $m_{0}$ this simplifies to $\omega
=2\bar{v}\kappa_{o}^{-1/2}\left(  3+\bar{t}t\right)  ^{-1}\kappa_{o}^{-1/2}v,$
which is then evaluated as $\omega=\ln\left(  27/16\right)  $ later in this
paper in Eqs.(\ref{intvv},\ref{intvv2}), to produce the result in
Eq.(\ref{g123}). Similarly, in the same computation, the relation between the
on-shell $g$ and the bare coupling $g_{0}$ initially takes the following form
$g_{0}=\frac{1}{2}g\left(  \frac{27}{16}\right)  ^{3/2}\det(\frac{3+m_{0}^{2}%
}{4})^{d/2}\det(\frac{3+m_{0}^{-2}}{4})^{-2/2}\left\vert \det m_{0}\right\vert
^{-d/4}\left\vert \det m_{0}^{-1}\right\vert ^{-2/4}.$ After inserting $m_{0}
$ it becomes
\begin{equation}
g_{0}=\frac{1}{2}g\left(  \frac{27}{16}\right)  ^{3/2}\left(  \det\left(
t\bar{t}\right)  \right)  ^{-\left(  d-6\right)  /4}\det\left(  \frac
{3+\left(  t\bar{t}\right)  }{4}\right)  ^{d}\det\left(  \frac{1+3\left(
t\bar{t}\right)  }{4}\right)  ^{-2},\;d=26. \label{g}%
\end{equation}
The right hand side of Eq.(\ref{g}) is divergent as $\left(  2N\right)
^{(d-6)/8-d/18+4/9}$ which becomes $\left(  2N\right)  ^{3/2}$ for $d=26$ as
will be later shown in Eq.(\ref{bare}). However this divergence is everywhere
reabsorbed into the definition of the on shell $g$ just as in the 3-tachyon
case. After this step, there still remains similar determinants that
individually produce a divergence or zeroes at large $N,$ but they combine
together to give finite answers magically as long as $d=26.$ We will see an
example of this impressive fact below in the 4-tachyon amplitude (see the
determinants in Eq.(\ref{a4}) and the computation in Eqs.(\ref{dets},)).

A similar computation can be performed for the vector particle for an incoming
wave with momentum $k^{\mu}$, which is represented by the string field
\begin{equation}
A_{V}\left(  x_{cm},x_{e},p_{e},k\right)  =e^{ik\cdot x_{cm}}A_{0}\left(
x_{e},p_{e}\right)  \varepsilon_{\mu}\left(  k\right)  \left(  p^{\mu
}T\right)  _{1}\sqrt{2\alpha^{\prime}},
\end{equation}
where $\varepsilon_{\mu}\left(  k\right)  $ is the polarization and the last
factor $p_{e}^{\mu}T_{e1}\sqrt{2\alpha^{\prime}}$ resulted from applying the
oscillator \cite{BM2}\cite{BKM3} $\alpha_{-1}^{\mu}A_{0}\left(  x_{e}%
,p_{e}\right)  $. For example, the coupling between the vector and two
tachyons is (omitting the Chan-Paton factors at the string ends)%
\begin{equation}
g_{12V_{3}}\left(  k_{i}\right)  =2g_{0}\int d^{d}\bar{x}Tr\left(
A_{0}e^{ik_{1}\cdot\left(  \bar{x}+wx_{e}\right)  }\star A_{0}e^{ik_{2}%
\cdot\left(  \bar{x}+wx_{e}\right)  }\star A_{0}e^{ik_{3}\cdot\left(  \bar
{x}+wx_{e}\right)  }\varepsilon_{\mu}\left(  k_{3}\right)  \left(  p^{\mu
}T\right)  _{1}\sqrt{2\alpha^{\prime}}\right)  .
\end{equation}
Following the monoid methods in \cite{BM2}\cite{BKM3} that led to
Eq.(\ref{g123}) this is evaluated in almost the same way, giving
\begin{equation}
g_{12V_{3}}\left(  k_{i}\right)  =\frac{1}{2}g\left(  k_{1}-k_{2}\right)
\cdot\varepsilon\left(  k_{3}\right)  \left(  \frac{27}{16}\right)
^{1-\frac{1}{2}\alpha^{\prime}\left(  k_{1}^{2}+k_{2}^{2}+k_{3}^{2}\right)  },
\end{equation}
with an implied momentum conservation delta function as in Eq.(\ref{g123}).
This result is in full agreement with the on-shell result in \cite{GJ} (for
$\alpha^{\prime}k_{1}^{2}=\alpha^{\prime}k_{2}^{2}=1$ and $\alpha^{\prime
}k_{3}^{2}=0$), as well as the off-shell results summarized in \cite{samuel}.
To arrive at this result we needed to perform the computation of the following
numerical coefficient
\begin{equation}
b_{1}=\sum_{e,e^{\prime}>0}T_{e1}\sqrt{e}\left(  \frac{2}{3+t\bar{t}}\right)
_{ee^{\prime}}\frac{1}{\sqrt{e^{\prime}}}w_{e^{\prime}}=\left(  \left(
\frac{2}{3+\bar{t}t}\right)  \kappa_{o}^{-1}v\right)  _{1}=\frac{2\sqrt{2}%
}{3\sqrt{3}}.
\end{equation}
This computation, and those for similar quantities will be given later in this
paper in Eq.(\ref{b1}).

We see that MSFT is in full agreement with previous computations in string
field theory for off-shell 3-point functions. Previous methods used conformal
maps \cite{samuel} and conformal field theory, while MSFT uses the Moyal
product, and produces the same results with considerably simpler methods. One
can go on computing very simply off-shell 3-point couplings for any other
perturbative or non-perturbative fields by similar MSFT techniques.

\section{Off-shell 4-tachyon scattering}

One of the new results in our paper is a proof that the MSFT 4-tachyon
amplitude produced by the Moyal product does give directly the Veneziano
amplitude, without using conformal mapping techniques. A second new result is
a comprehensive understanding of the off-shell factor by going considerably
further in the parametric expansion, obtaining a plot of the function in the
full range, and determining its differentiability at a turning point.

In MSFT we have been developing analytic methods of computation in string
field theory by using directly the Moyal star product. Both the oscillator
formalism and MSFT provide an expression for any off-shell amplitude,
including loops. Although the starting point and intermediate steps are quite
different, it has been argued that generally we expect agreement in the result
\cite{BKM1}\cite{BKM3}.

The computation of the 4-tachyon off-shell amplitude in MSFT was performed in
\cite{BKM1}\cite{BKM3}. There are 8 diagrams that correspond to various
permutations of the four external legs. The $s$-channel diagram $_{1}^{2}%
{>}-{<}_{4}^{3}$ is denoted as $_{12}A_{34}.$ The mathematical expression for
this amplitude was obtained in \cite{BKM1}\cite{BKM3}. The $t$-channel diagram
is given by the cyclic permutation of the external legs $_{41}A_{23},$ and its
mathematical expression amounts to exchanging $s,t$ in the previous result.
The remaining permutations that do not change the $s,t$ channel properties are
denoted as $_{21}A_{43},$ $_{43}A_{21}$, $_{34}A_{12}$ and $_{14}A_{32},$
$_{23}A_{41},$ $_{32}A_{14}$ respectively. After a brief computation (as in
\cite{BKM1}) it can be shown that these give the same amplitudes as the
initial $s,t$ diagrams. Therefore the sum of the diagrams for the $s,t$
channels produces a factor of 4
\begin{align}
A\left(  s,t\right)   &  =4\left(  _{1}^{2}{>}-{<}_{4}^{3}+%
\genfrac{.}{.}{0pt}{}{^{2}\underset{_{|}}{\vee}^{3}}{_{1}\wedge_{4}}%
\right)  =4\int_{0}^{\infty}d\tau e^{\tau}~\left(  ~_{12}A_{34}\left(
\tau\right)  +~~_{41}A_{{\small 23}}\left(  \tau\right)  \right)  \nonumber\\
&  =\int_{0}^{\infty}d\tau~e^{-\tau-2\alpha(\tau)+4\gamma(\tau)}a_{4}\left(
\tau\right)  ~e^{\left(  \gamma(\tau)+2\beta\left(  \tau\right)  \right)
\sum_{i=1}^{4}\left(  \alpha^{\prime}k_{i}^{2}-1\right)  }\label{Atau}\\
&  \times\left[  e^{\left(  \alpha^{\prime}s+2\right)  (\tau+\alpha
(\tau)+2\beta(\tau))}~e^{\left(  \alpha^{\prime}t+2\right)  2\beta(\tau
)}+\left(  s\leftrightarrow t\right)  \right]  .\nonumber
\end{align}
where $s,t$ are the Mandelstam variables\footnote{The Mandelstam $t$ should
not ne confused with the matrix $t_{eo}.$} $s=-\left(  k_{1}+k_{2}\right)
^{2}$ and $t=-\left(  k_{1}+k_{4}\right)  ^{2}$. The functions $\alpha\left(
\tau\right)  ,\beta\left(  \tau\right)  ,\gamma\left(  \tau\right)
,a_{4}\left(  \tau\right)  $ initially given in \cite{BKM1} were further
simplified in \cite{BKM3}. Here, as a first step, we simplify these functions
further with a few algebraic steps and put them into the following form which
will be convenient in our analysis
\begin{align}
\alpha\left(  \tau\right)   &  =\omega-\bar{b}e^{-\tau\kappa}\left(
1-Me^{-\tau\kappa}\right)  ^{-1}b,\;\;\label{alpha}\\
\beta\left(  \tau\right)   &  =\bar{b}P^{\left(  o\right)  }e^{-\tau\kappa
}\left(  1-Me^{-\tau\kappa}Me^{-\tau\kappa}\right)  ^{-1}P^{\left(  o\right)
}b,\label{beta}\\
\gamma\left(  \tau\right)   &  =-\frac{1}{2}\omega-\bar{b}P^{\left(  o\right)
}e^{-\tau\kappa}\left(  1-Me^{-\tau\kappa}\right)  ^{-1}P^{\left(  o\right)
}b,\label{gamma}\\
a_{4}\left(  \tau\right)   &  =4\frac{1}{4}g^{2}\left(  \frac{27}{16}\right)
^{3}\frac{\det\left(  1-\tilde{M}e^{-\tau\kappa}\tilde{M}e^{-\tau\kappa
}\right)  }{\left[  \det\left(  1-Me^{-\tau\kappa}Me^{-\tau\kappa}\right)
\right]  ^{d/2}},\;d=26.\label{a4}%
\end{align}
The overall factor of 4 in $a_{4}$ is the factor in Eq.(\ref{Atau}). In string
mode space labelled by even and odd positive integers $e=2,4,6,\cdots$ and
$o=1,3,5,\cdots,$ we have defined the matrices $\kappa,M,\tilde{M},P^{\left(
o\right)  },$ vector $b$ and scalar $\omega$ as follows (a bar above a symbol
means matrix transpose)%
\begin{align}
\kappa &  =\left(
\begin{array}
[c]{cc}%
e & 0\\
0 & o
\end{array}
\right)  ,\;M=\left(
\begin{array}
[c]{cc}%
\left(  \frac{1-t\bar{t}}{3+t\bar{t}}\right)  _{ee^{\prime}} & 0\\
0 & \left(  \frac{1-\bar{t}t}{3+\bar{t}t}\right)  _{oo^{\prime}}%
\end{array}
\right)  ,\;\tilde{M}=\left(
\begin{array}
[c]{cc}%
\left(  \frac{1-t\bar{t}}{1+3t\bar{t}}\right)  _{ee^{\prime}} & 0\\
0 & \left(  \frac{1-\bar{t}t}{1+3\bar{t}t}\right)  _{oo^{\prime}}%
\end{array}
\right)  \label{kappa}\\
b &  =\left(
\begin{array}
[c]{c}%
b_{e}\\
b_{o}%
\end{array}
\right)  ,\;P^{\left(  o\right)  }=\left(
\begin{array}
[c]{cc}%
0 & 0\\
0 & 1
\end{array}
\right)  ,\;\omega\equiv\bar{b}\left(  1-M\right)  ^{-1}b.\label{b}%
\end{align}
Here the diagonal matrix $\kappa$ with integer eigenvalues $\kappa_{n}=n,$
represents the spectrum of string oscillation frequencies (eigenvalues of
$L_{0}$), $P^{\left(  o\right)  }$ is a projector onto the odd modes only. The
other quantities are all constructed from the matrix $t_{eo}$ and vector
$v_{o}$ which themselves are built from the frequencies $\kappa_{n}$ as given
in \cite{B1}-\cite{BKM1} for any $N$. In the large $N$ limit they take the
form in Eq.(\ref{v}). The combination of $t_{eo}$ in the form of the matrix
$M$ arises frequently in the interactions in the matter sector, while the
matrix $\tilde{M}$ occurs in the ghost sector\footnote{The matrices
$M,\tilde{M}$ is are simplified forms of $-\mathcal{M}^{\left(  0\right)
},-CX^{\left(  0\right)  }$ where $\mathcal{M}^{\left(  0\right)  },X^{\left(
0\right)  }$ were identified in \cite{BM2}\cite{BKM3} as some of the Neumann
matrices for the 3-point vertex in the matter and ghost sectors respectively.
Furthermore, it was shown that all Neumann matrices, for all $n$-point
vertices, are explicit functions of the matrix $t_{eo}$ as obtained in
\cite{BM2}\cite{BKM3}. Hence $t$ is the fundamental matrix that determines all
interactions in string theory.}. The even and odd vectors $b_{e},b_{o}$ are
given by%
\begin{equation}
b_{o}=\left(  \frac{2}{3+\bar{t}t}\kappa_{o}^{-1/2}v\right)  _{o}%
,\;b_{e}=\left(  t\frac{2}{3+t\bar{t}}\kappa_{o}^{-1/2}v\right)
_{e},\label{bobe}%
\end{equation}
where $\kappa_{o}$ is the odd part of the matrix $\kappa.$ Finally $\omega,$
which is the $\tau\rightarrow\infty$ limit of $\alpha\left(  \tau\right)  $ is
simplified to the form%
\begin{equation}
\omega\equiv\bar{b}\left(  1-M\right)  ^{-1}b=\bar{v}\kappa_{o}^{-1}\frac
{2}{3+\bar{t}t}\kappa_{o}^{-1}v\label{alinf}%
\end{equation}
by using the definitions of $b,M$ given above.

From these quantities we compute the functions $\alpha\left(  \tau\right)
,\beta\left(  \tau\right)  ,\gamma\left(  \tau\right)  ,a_{4}\left(
\tau\right)  $ which in turn determine the 4-tachyon scattering amplitude
off-shell as well as on shell. We will show that the on-shell amplitude for
$\alpha^{\prime}k_{i}^{2}=1$ reduces to the Veneziano amplitude given by the
beta function
\begin{equation}
A_{shell}\left(  s,t\right)  =g^{2}\frac{\Gamma\left(  -\alpha^{\prime
}s-1\right)  \Gamma\left(  -\alpha^{\prime}t-1\right)  }{\Gamma\left(
-\alpha^{\prime}s-\alpha^{\prime}t-2\right)  }. \label{venez}%
\end{equation}
On the other hand the off-shell expression above goes beyond conformal field
theory which only gives information for on-shell strings.

The aim in the rest of this section is to show that the off-shell 4-tachyon
amplitude in Eq.(\ref{Atau}) can be rewritten in the following form and then
compute the function $f\left(  x\right)  $%
\begin{align}
A\left(  s,t\right)   &  =-g^{2}\int_{0}^{\infty}d\tau\frac{dx\left(
\tau\right)  }{d\tau}\left(  f\left(  x\left(  \tau\right)  \right)  \right)
^{\sum_{i=1}^{4}\left(  \alpha^{\prime}k_{i}^{2}-1\right)  }\label{Axtau}\\
&  \times\left[  \left(  x\left(  \tau\right)  \right)  ^{-\alpha^{\prime}%
s-2}\left(  1-x\left(  \tau\right)  \right)  ^{-\alpha^{\prime}t-2}+\left(
s\leftrightarrow t\right)  \right] \nonumber\\
&  =g^{2}\int_{0}^{1}dx~x^{-\alpha^{\prime}s-2}\left(  1-x\right)
^{-\alpha^{\prime}t-2}\left(  f\left(  x\right)  \right)  ^{\sum_{i=1}%
^{4}\left(  \alpha^{\prime}k_{i}^{2}-1\right)  }. \label{Ax}%
\end{align}
After the change of integration variables from $\tau$ to $x$ in the form of
Eq.(\ref{Ax}) we see that the off-shell amplitude is consistent with the
on-shell Veneziano amplitude (beta function in Eq.(\ref{venez})) when
$\alpha^{\prime}k_{i}^{2}=1$.

We will show that the change of variables from $\tau$ to $x$ is such that
$x\left(  0\right)  =1/2$ and $x\left(  \infty\right)  =0.$ Then we see that
the $s$-channel amplitude $_{12}A_{{\small 34}}\left(  \tau\right)  =_{1}^{2}$%
$>$%
$--$%
$<$%
$_{4}^{3}$ contributes to the range $0\leq x\leq1/2$ while the $t$-channel
amplitude $_{41}A_{{\small 23}}\left(  \tau\right)  $ contributes to the range
$1/2\leq x\leq1$ after a change of variables $x\rightarrow\left(  1-x\right)
.$ For the first form in Eq.(\ref{Atau}) to agree with the second form in
Eq.(\ref{Axtau}) it is required that $\alpha\left(  \tau\right)  ,\beta\left(
\tau\right)  ,\gamma\left(  \tau\right)  ,a_{4}\left(  \tau\right)  $ conspire
to have remarkable relations among themselves so that they can be written as
functions of the same $x\left(  \tau\right)  .$ Thus we need to prove that the
following relations are satisfied (which also define $x\left(  \tau\right)  $
in terms of $\alpha\left(  \tau\right)  ,\beta\left(  \tau\right)
,\gamma\left(  \tau\right)  ,a_{4}\left(  \tau\right)  $)
\begin{align}
e^{-\tau}e^{-\alpha\left(  \tau\right)  -2\beta\left(  \tau\right)  }  &
=1-e^{-2\beta\left(  \tau\right)  }\equiv x\left(  \tau\right)
,\label{xoftau}\\
\frac{dx\left(  \tau\right)  }{d\tau}  &  =-\frac{a_{4}\left(  \tau\right)
}{g^{2}}e^{-\tau-2\alpha(\tau)+4\gamma(\tau)}. \label{dx}%
\end{align}
Note that $a_{4}\left(  \tau\right)  $ depends on $d=26$ while the other
quantities are independent of the number of dimensions. Hence if the relation
holds for $d=26$ it cannot hold for other dimensions. We will prove below that
the relations in Eqs.(\ref{xoftau},\ref{dx}) are indeed true. This makes
$d=26$ unique.

Once the relations are proven, then we learn that the off-shell factor
$f\left(  x\right)  $ is given by%
\begin{equation}
f\left(  x\left(  \tau\right)  \right)  =e^{\gamma(\tau)+2\beta\left(
\tau\right)  }.
\end{equation}
To write $f\left(  x\right)  $ in terms of only the parameter $x$ in the
integral representation of the off-shell amplitude in Eq.(\ref{Ax}), the
relation between $x,\tau$ given in Eq.(\ref{xoftau}) needs to be inverted
$\tau=\tau\left(  x\right)  $. We will perform the inversion and will
construct the function $f\left(  x\right)  $ as a series expansion in powers
of $x.$ It turns out that a few terms in the expansion already give the
necessary information to obtain a sufficiently accurate representation of the
function $f\left(  x\right)  ,$ and hence of the full off-shell 4-tachyon amplitude.

Let us first prove that the relations in Eqs.(\ref{xoftau}) hold at the
integration limits $\tau=0,\infty.$ In the next section we show how to compute
the functions $\alpha\left(  \tau\right)  ,\beta\left(  \tau\right)
,\gamma\left(  \tau\right)  ,a_{4}\left(  \tau\right)  $ at $\tau=0,\infty.$
We find in particular%
\begin{align}
\alpha\left(  0\right)   &  =0,\;\beta\left(  0\right)  =\frac{1}{2}%
\ln2,\;\gamma\left(  0\right)  =-\ln2,\;\label{zero}\\
\alpha\left(  \infty\right)   &  =\ln\frac{27}{16}\allowbreak,\;\beta\left(
\infty\right)  =0,\;\gamma\left(  \infty\right)  =\frac{1}{2}\ln\frac{16}%
{27},\;a_{4}\left(  \infty\right)  =g^{2}\left(  \frac{27}{16}\right)  ^{3}.
\label{inf}%
\end{align}
From this we see that indeed Eqs.(\ref{xoftau}) is satisfied at both limits
$\tau=0,\infty$, and we also determine%
\begin{equation}
x\left(  0\right)  =\frac{1}{2},\;x\left(  \infty\right)  =0. \label{limits}%
\end{equation}
This shows that the $s$-channel amplitude $_{12}A_{{\small 34}}\left(
\tau\right)  $ is associated with the range $0\leq x\leq1/2$ and the
$t$-channel amplitude $_{41}A_{{\small 23}}\left(  \tau\right)  $ contributes
to the range $1/2\leq x\leq1$ after the change of variable $x\rightarrow
\left(  1-x\right)  .$

Next we examine the relations for more general values of $\tau.$ The form of
the equations in Eqs.(\ref{alpha}-\ref{a4}) was developed to examine an
expansion in powers of $e^{-\tau}$. In the next section we show how to compute
the coefficients for the large $\tau$ expansion of the functions
$\alpha\left(  \tau\right)  ,\beta\left(  \tau\right)  ,\gamma\left(
\tau\right)  ,a_{4}\left(  \tau\right)  .$ We find the following analytic
result
\begin{align}
\alpha\left(  \tau\right)   &  =-\ln\frac{16}{27}\allowbreak-\frac{8}%
{27}e^{-\tau}-\frac{2^{2}19}{\left(  27\right)  ^{2}}e^{-2\tau}-\frac
{2^{5}7^{2}}{3\left(  27\right)  ^{3}}e^{-3\tau}\nonumber\\
&  -\frac{2\times13^{2}43}{\left(  27\right)  ^{4}}e^{-4\tau}-\frac
{2^{4}167\times229}{5\left(  27\right)  ^{5}}\allowbreak e^{-5\tau}+O\left(
e^{-6\tau}\right)  ,\\
\beta\left(  \tau\right)   &  =\frac{8}{27}e^{-\tau}+\frac{2^{5}7^{2}%
}{3\left(  27\right)  ^{3}}e^{-3\tau}+\frac{2^{4}167\times229}{5\left(
27\right)  ^{5}}e^{-5\tau}+O\left(  e^{-7\tau}\right)  ,\\
\gamma\left(  \tau\right)   &  =\frac{1}{2}\ln\frac{16}{27}-\frac{8}%
{27}e^{-\tau}-\frac{40}{\left(  27\right)  ^{2}}e^{-2\tau}-\frac{2^{5}7^{2}%
}{3\left(  27\right)  ^{3}}e^{-3\tau}\nonumber\\
&  -\frac{2^{3}829}{\left(  27\right)  ^{4}}\allowbreak e^{-4\tau}-\frac
{2^{4}167\times229}{5\left(  27\right)  ^{5}}\allowbreak e^{-5\tau}+O\left(
e^{-6\tau}\right)  ,\\
a_{4}\left(  \tau\right)   &  =g^{2}\left(  \frac{27}{16}\right)  ^{3}\left(
1+\frac{2^{2}17}{3^{5}}e^{-2\tau}+\frac{2\times1399}{\left(  27\right)  ^{3}%
}e^{-4\tau}+O\left(  z^{6}\right)  \right)  .
\end{align}
To obtain the expansion to this order it is sufficient to compute the
coefficients $b_{1},b_{2},b_{3},b_{4},b_{5},M_{11},M_{13},M_{22},\tilde
{M}_{11},\tilde{M}_{13},\tilde{M}_{22}$ defined above. These results were
obtained analytically without much effort. Our analytic results above to order
$O\left(  e^{-6\tau}\right)  $ are already quite adequate to construct
$f\left(  x\right)  .$ It is possible to easily extend the expansion by
inserting the results for the higher coefficients provided in the appendix
into Eqs.(\ref{alpha}-\ref{a4}). In fact we have constructed an algebraic
computer program that does this, and using it we have double checked our
analytic results above and extended our computation to higher orders. We will
report on some of the higher order results below.

The relations (\ref{xoftau},\ref{dx}) can now be verified directly by
inserting the large $\tau$ expansions for $\alpha\left(  \tau\right)
,\beta\left(  \tau\right)  ,\gamma\left(  \tau\right)  ,a_{4}\left(
\tau\right)  $ given above, and re-expanding in powers of $e^{-\tau}$ up to
order $O\left(  e^{-7\tau}\right)  $. From either the first or second term in
Eq.(\ref{xoftau}) we obtain the same expression for $x\left(  \tau\right)  ,$
namely%
\begin{align}
x\left(  \tau\right)   &  =\left(
\begin{array}
[c]{c}%
\allowbreak\frac{16}{27}e^{-\tau}-\frac{128}{729}e^{-2\tau}+\frac{64}%
{729}e^{-3\tau}-\frac{19\,456}{531\,441}e^{-4\tau}\\
+\frac{387\,296}{14\,348\,907}\allowbreak e^{-5\tau}-\frac{1733\,120}%
{129\,140\,163}e^{-6\tau}+O\left(  e^{-7\tau}\right)
\end{array}
\right)  ,\label{xExp}\\
&  =\left(
\begin{array}
[c]{c}%
59.\,\allowbreak259e^{-\tau}-17.\,\allowbreak558e^{-2\tau}+\allowbreak
8.\,\allowbreak779\,1e^{-3\tau}-3.\,\allowbreak661\,0e^{-4\tau}\\
+\allowbreak2.\,\allowbreak699\,1e^{-5\tau}-1.\,\allowbreak342e^{-6\tau
}+O\left(  e^{-7\tau}\right)
\end{array}
\right)  \times10^{-2}\label{xExp2}%
\end{align}
Furthermore, the right hand side of Eq.(\ref{dx}) gives precisely the
derivative $\partial_{\tau}x\left(  \tau\right)  $ of the expansion in
Eq.(\ref{xExp})%
\begin{equation}
-\frac{a_{4}\left(  \tau\right)  }{g_{T}^{2}}e^{-\tau-2\alpha(\tau
)+4\gamma(\tau)}=\left(
\begin{array}
[c]{c}%
-\frac{16}{27}e^{-\tau}+\frac{256}{729}e^{-2\tau}-\frac{64}{243}e^{-3\tau
}+\frac{77\,824}{531\,441}\allowbreak e^{-4\tau}\\
-\frac{1936\,480}{14\,348\,907}e^{-5\tau}+\frac{3466\,240}{43\,046\,721}%
e^{-6\tau}+O\left(  e^{-7\tau}\right)
\end{array}
\right)  .\allowbreak
\end{equation}
This proves that $\alpha\left(  \tau\right)  ,\beta\left(  \tau\right)  $
satisfy the relation in Eq.(\ref{xoftau}), and $\alpha\left(  \tau\right)
,\gamma\left(  \tau\right)  ,a_{4}\left(  \tau\right)  $ satisfy the relation
in Eq.(\ref{dx}), at least up to order $O\left(  e^{-7\tau}\right)  .$
Actually, as already mentioned, with a computer program we have shown that the
relations hold to much higher orders. These results are convincing that
$\alpha\left(  \tau\right)  ,\beta\left(  \tau\right)  ,\gamma\left(
\tau\right)  ,a_{4}\left(  \tau\right)  $ are all expressed in terms of the
same function $x\left(  \tau\right)  .$ We emphasize that the number of
dimensions $d$ which appears in $a_{4}\left(  \tau\right)  $ in Eq.(\ref{a4})
must be $d=26$ to satisfy the relations.

Note that at $\tau=0$ we expect $x\left(  0\right)  =1/2$ exactly as in
Eq.(\ref{limits}), and indeed by replacing $e^{-\tau}\rightarrow1$ the
expansion (\ref{xExp}) we obtain $x\left(  0\right)  =\allowbreak0.481\,76$
which implies that the expansion captures an accurate representation of the
full function. Note from the trend in Eq.(\ref{xExp2})\ that the next
$O\left(  e^{-7\tau}\right)  $ correction will bring the value much closer to
the exact answer $x\left(  0\right)  =1/2$. With the computer program we have
shown that this value becomes $x\left(  0\right)  =0.492\,95$ by computing
$x\left(  \tau\right)  $ to order $O\left(  e^{-10\tau}\right)  $.

Finally we compute the off-shell factor, which is given by
\begin{equation}
f\left(  x\left(  \tau\right)  \right)  =e^{\gamma(\tau)+2\beta\left(
\tau\right)  }\equiv\frac{C\left(  1-x\left(  \tau\right)  \right)  }%
{2\sqrt{1-x\left(  \tau\right)  }}\label{fxt}%
\end{equation}
The expression on the right hand side defines the function $C$ which is given
here for comparison to the old literature. Thus, by substituting $1-x\left(
\tau\right)  =\exp\left(  -2\beta\left(  \tau\right)  \right)  $ we can write
$C\left(  1-x\left(  \tau\right)  \right)  =2e^{\gamma(\tau)+\beta\left(
\tau\right)  }.$ First we obtain the expansion of $C\left(  1-x\left(
\tau\right)  \right)  $ by inserting our computation of $\gamma\left(
\tau\right)  $ and $\beta\left(  \tau\right)  .$ The result is
\begin{equation}
C\left(  1-x\left(  \tau\right)  \right)  =2e^{\gamma(\tau)+\beta\left(
\tau\right)  }=\left(  \frac{4}{3}\right)  ^{3/2}\left(  1-\frac{5}{32}\left(
\frac{16}{27}\right)  ^{2}e^{-2\tau}-\frac{1}{32}\left(  \frac{16}{27}\right)
^{2}e^{-4\tau}+O\left(  e^{-6\tau}\right)  \right)
\end{equation}
Next, by using Eq.(\ref{xExp}) we rewrite this result in terms of $x\left(
\tau\right)  $ so that the function $C\left(  1-x\left(  \tau\right)  \right)
$ is given in terms of $x.$ The result is%
\begin{align}
C\left(  1-x\left(  \tau\right)  \right)   &  =\left(  \frac{4}{3}\right)
^{3/2}\left(
\begin{array}
[c]{c}%
1-\frac{5}{32}\left(  x\left(  \tau\right)  \right)  ^{2}-\frac{5}{32}\left(
x\left(  \tau\right)  \right)  ^{3}\\
-\frac{1249}{8192}\left(  x\left(  \tau\right)  \right)  ^{4}-\frac{609}%
{4096}\left(  x\left(  \tau\right)  \right)  ^{5}+O\left(  \left(  x\left(
\tau\right)  \right)  ^{6}\right)
\end{array}
\right)  \\
&  =\left(
\begin{array}
[c]{c}%
1.\,\allowbreak539\,6-0.240\,56x^{2}-\allowbreak0.240\,56x^{3}\\
-0.234\,74\allowbreak x^{4}-0.228\,91x^{5}+O\left(  x^{6}\right)
\end{array}
\right)  \nonumber
\end{align}
These coefficients are determined by demanding that the expansion in powers of
$e^{-\tau}$ matches the one of $2e^{\gamma(\tau)+\beta\left(  \tau\right)  }.$
In this form our result for $C$ is in agreement with what is found in the old
literature \cite{sloan} where $C$ was computed up to the second term
$-\frac{5}{32}x^{2}$ by using Mandelstam's conformal mapping techniques. We,
of course, used the very different Moyal star technique and obtained the same
result, but we also easily went further by obtaining the higher order terms,
which is a new result given in this paper. As mentioned, with our technique it
is very easy to compute to even higher orders (see Eq.(\ref{expansion})).

Since the range for $x$ is $0\leq x$ $\leq1/2$ the expansion given above for
$C$ has good convergence, so we expect that we have obtained an accurate
representation of the full function $C$ in the relevant range. As a test let
us compare the exact value of $C$ at $\tau=0$ (or at $x=1/2$) which we can
compute exactly $C\left(  1-x\left(  0\right)  \right)  =2e^{\gamma
(0)+\beta\left(  0\right)  }=\sqrt{2}=\allowbreak1.\,\allowbreak414\,2$ after
using Eq.(\ref{zero})$.$ By evaluating the expansion above at $x=1/2,$ we
obtain $C=1.\,\allowbreak427\,6,$ which confirms that the expansion does
capture the function almost fully in the entire range.

Let us now turn to the full off-shell factor $f\left(  x\right)  .$ The exact
expansion of this function in powers of $x$ up to $\allowbreak O\left(
x^{6}\right)  $ becomes%
\begin{align}
f\left(  x\right)   &  =\frac{C\left(  1-x\right)  }{2\left(  1-x\right)
^{1/2}}=\frac{4}{3\sqrt{3}}\left(
\begin{array}
[c]{c}%
1+\frac{1}{2}x+\frac{7}{32}x^{2}+\frac{5}{64}x^{3}-\frac{129}{8192}x^{4}\\
-\frac{1413}{16\,384}x^{5}+\allowbreak O\left(  x^{6}\right)
\end{array}
\right) \label{series}\\
&  =\frac{4}{3\sqrt{3}}\left(
\begin{array}
[c]{c}%
1+0.5x+0.218\,75\allowbreak x^{2}+0.07813x^{3}\\
-0.01575\allowbreak x^{4}-0.08624x^{5}+\allowbreak O\left(  x^{6}\right)
\end{array}
\right)
\end{align}
At the end of the range $x=1/2$ (i.e. $\tau=0$) the exact value of this
function should be $1,$ since%
\begin{equation}
f\left(  1/2\right)  =f\left(  x\left(  0\right)  \right)  =e^{\gamma
(0)+2\beta\left(  0\right)  }=1, \label{peak}%
\end{equation}
where we used Eq.(\ref{zero}). The series approximation in powers of $x$ given
above to order $O\left(  x^{5}\right)  $ produces $f\left(  1/2\right)
=1.\,\allowbreak0090,$ which is better than $1\%$ accuracy.

To get a better feel of the function $f\left(  x\right)  $ we plot the
expansion in Eq.(\ref{series}) for the appropriate range for $x$ which is
$0\leq x$ $\leq1/2.$ This is shown by the solid line in Fig.1.%
\begin{center}
\includegraphics[
height=1.9969in,
width=2.9957in
]%
{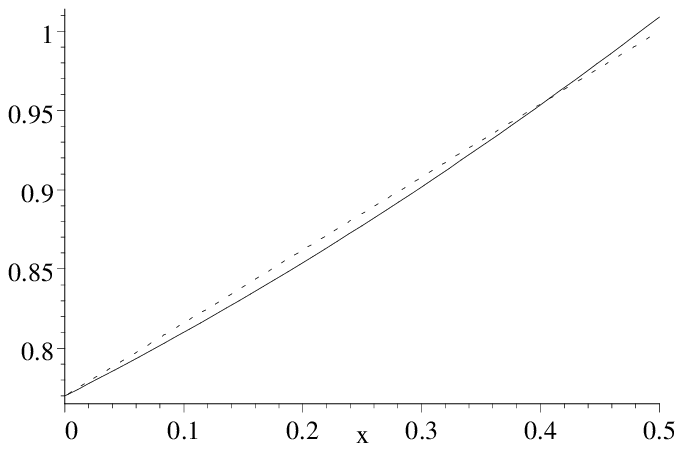}%
\\
Fig.1
\end{center}
The figure suggests that, for the relevant range, as a guide for the eye we
can compare $f\left(  x\right)  $ roughly to a linear function $\tilde
{f}\left(  x\right)  $
\begin{equation}
\tilde{f}\left(  x\right)  =\frac{4}{3\sqrt{3}}+2\left(  1-\frac{4}{3\sqrt{3}%
}\right)  x,
\end{equation}
where the slope is chosen to guarantee the exact values at both ends of the
range, namely $\frac{4}{3\sqrt{3}}$ at $x=0$ and $1$ at $x=1/2.$ This case
corresponds to the dotted line in the figure. The plots suggest that the exact
curve $f\left(  x\right)  $ lies somewhere close to the solid line and the
dotted line, and that in any case we have obtained a fairly good approximation
in the entire range with the series expansion in Eq.(\ref{series}).

By combining the $s$ and $t$ channel contributions we extend the range of
integration to $0\leq x\leq1$ as explained above. Thus, in the range $1/2\leq
x\leq1$ the function $f\left(  x\right)  $ is given by substituting $x$ by
$\left(  1-x\right)  $ in the expression of $f\left(  x\right)  $ given above.
Namely, the plot of $f\left(  x\right)  $ in the full range $0\leq x\leq1$ is
given in Fig.2.
\begin{center}
\includegraphics[
height=1.9969in,
width=2.9957in
]%
{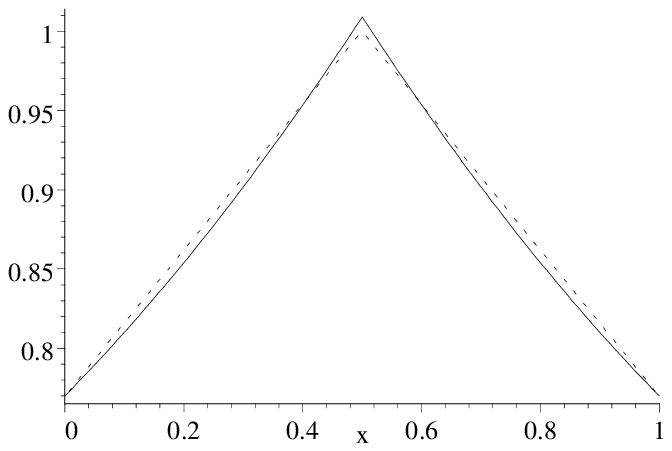}%
\\
Fig.3
\end{center}
The exact function is expected to lie somewhere close to the solid and the
dotted lines. The shape of the curve begs the question of whether the exact
$f\left(  x\right)  $ ever crosses the dotted line. To provide an answer we
used our algebraic computer program in which we plugged in the contents of the
Appendix into Eqs.(\ref{alpha}-\ref{a4}) and obtained the following higher
order expansion (we do not give the details of the expansion for
$\alpha\left(  \tau\right)  ,\beta\left(  \tau\right)  ,\gamma\left(
\tau\right)  ,a_{4}\left(  \tau\right)  $)
\begin{equation}
f\left(  x\right)  =\frac{4}{3\sqrt{3}}\left(
\begin{array}
[c]{c}%
1+\frac{1}{2}x+\frac{7}{32}x^{2}+\frac{5}{64}x^{3}-\frac{129}{8192}x^{4}%
-\frac{1413}{16\,384}x^{5}-\frac{40\,973}{262\,144}x^{6}\\
-\frac{124\,459}{524\,288}x^{7}-\frac{186\,841\,777}{536\,870\,912}x^{8}%
-\frac{547\,864\,633}{1073\,741\,824}x^{9}+O\left(  x^{10}\right)
\end{array}
\right)  .\label{expansion}%
\end{equation}
$\allowbreak$The plot of this more accurate expansion lies slightly below the
solid line in the figure, and is extremely close to it in the regions that are
not near to $x=1/2.$ It does cross the dotted line, but it does it at mirror
points closer to $x=1/2$ compared to Fig.2, and finally reaches the value
$f\left(  1/2\right)  =\allowbreak1.\,\allowbreak003\,9$ at $x=1/2.$ So the
new corrected peak is between the previous value of $1.0090$ and the exact
value $1.0090>1.\,\allowbreak003\,9>1$. This analysis is consistent with the
possibility that the exact function $f\left(  x\right)  $ may cross the dotted
line somewhere close to $x=1/2$ before settling into the value $f\left(
1/2\right)  =1.$ This issue of crossing or not crossing is relevant for
determining whether the exact function is differentiable at $x=1/2.$ If it
never crosses the dotted line then it must make a cusp and be
non-differentiable. Such a non-differentiable function may be a peculiarity of
Witten's theory in which a particular conformal gauge has been effectively
fixed by distinguishing the midpoint. We remind the reader that off-shell
amplitudes generally are not gauge invariant.

It seems therefore interesting to investigate the slope of $f\left(  x\right)
$ near $x=1/2.$ A glimpse of the slope on the left side of the possible cusp
is obtained by computing the derivative of the expansion above and evaluating
it at $x=1/2.$ We obtain $\partial_{x}f\left(  x\right)  |_{x=1/2}%
\simeq0.49849$ consistent with a cusp. However it is not clear how much we can
trust this number, because if we examine the size of the coefficients in the
expansion of the derivative%
\begin{equation}
\partial_{x}f\left(  x\right)  =\left(
\begin{array}
[c]{c}%
38.\,\allowbreak49+16.\,\allowbreak839\left(  2x\right)  +4.\,\allowbreak
510\,5\allowbreak\left(  2x\right)  ^{2}-0.606\,11\left(  2x\right)
^{3}-2.\,\allowbreak074\,7\allowbreak\left(  2x\right)  ^{4}\\
-2.\,\allowbreak256\,0\left(  2x\right)  ^{5}-1.\,\allowbreak998\,7\left(
2x\right)  ^{6}-1.\,\allowbreak674\,4\left(  2x\right)  ^{7}-1.\,\allowbreak
380\,9\allowbreak\left(  2x\right)  ^{8}+O\left(  x^{9}\right)
\end{array}
\right)  \times10^{-2}%
\end{equation}
we see that convergence is not very fast, and the exact value may turn out to
be quite different at $2x=1$ since there are an infinite number of terms to be
summed. Note that the tendency of the neglected terms is to be negative
thereby reducing the value of the slope given above. To settle the question we
must examine an expansion near $\tau=0$ rather than near $\tau=\infty,$ but
the reason we did not do this so far in this paper is because the expansion
does not seem to exist due to divergent derivatives near $\tau=0$. However,
precisely these divergences turn out to settle the issue as follows.

We can compute the derivative $\left(  \partial_{x}f\left(  x\right)  \right)
_{x=1/2}$ from $\partial_{\tau}f\left(  x\left(  \tau\right)  \right)
|_{\tau=0}$ as follows%
\begin{equation}
\dot{x}\left(  0\right)  \left(  \partial_{x}f\left(  x\right)  \right)
_{x=1/2}=\partial_{\tau}f\left(  x\left(  \tau\right)  \right)  |_{\tau
=0}=\left(  \dot{\gamma}\left(  0\right)  +2\dot{\beta}\left(  0\right)
\right)  e^{\gamma\left(  0\right)  +2\beta\left(  0\right)  },
\end{equation}
where we have used the chain rule on the left hand side and Eq.(\ref{fxt}) on
the right hand side. From Eq.(\ref{xoftau}) we have $\dot{x}\left(  0\right)
=2\dot{\beta}\left(  0\right)  e^{-2\beta\left(  0\right)  }.$ Inserting the
exact values for $\gamma\left(  0\right)  ,\beta\left(  0\right)  $ given in
Eq.(\ref{zero}) these equations become%
\begin{equation}
\dot{x}\left(  0\right)  =\dot{\beta}\left(  0\right)  ,\;\;\dot{x}\left(
0\right)  \left(  \partial_{x}f\left(  x\right)  \right)  _{x=1/2}=\dot
{\gamma}\left(  0\right)  +2\dot{\beta}\left(  0\right)  . \label{slope}%
\end{equation}
By differentiating Eqs.(\ref{beta},\ref{gamma}) we obtain expressions for
$\dot{\gamma}\left(  0\right)  ,\dot{\beta}\left(  0\right)  $%
\begin{align}
\dot{\beta}\left(  0\right)   &  =-\bar{b}_{o}\left(  1-M_{o}^{2}\right)
^{-1}\left(  \kappa_{o}+M_{o}\kappa_{o}M_{o}\right)  \left(  1-M_{o}%
^{2}\right)  ^{-1}b_{o},\\
\dot{\gamma}\left(  0\right)   &  =\bar{b}_{o}\left(  1-M_{o}\right)
^{-1}\kappa_{o}\left(  1-M_{o}\right)  ^{-1}b_{o},
\end{align}
where $M_{o}=\frac{1-\bar{t}t}{3+\bar{t}t}$ is the odd part of the matrix $M.
$ After srtaightforward algebra, we find%
\begin{equation}
\dot{\gamma}\left(  0\right)  =\bar{v}\kappa_{o}^{-1/2}\frac{1}{\left(
1+\bar{t}t\right)  }\kappa_{o}\frac{1}{\left(  1+\bar{t}t\right)  }\kappa
_{o}^{-1/2}v, \label{gammadot}%
\end{equation}
and $\dot{x}\left(  0\right)  ,\dot{\beta}\left(  0\right)  $ related to the
same expression by%
\begin{equation}
\dot{x}\left(  0\right)  =\dot{\beta}\left(  0\right)  =-\frac{1}{2}%
\dot{\gamma}\left(  0\right)  -\frac{1}{2}.
\end{equation}
We used $\bar{v}v=1$ which multiplied the last constant term $-1/2.$ In
passing we mention that we have also computed $\dot{\alpha}\left(  0\right)  $
and found $\dot{\alpha}\left(  0\right)  =2\dot{\gamma}\left(  0\right)  $
after a little algebra. Through Eqs.(\ref{vfkfv},\ref{VkoV}) and
Eqs.(\ref{klok}-\ref{klk}) in the next section we show that the expression
above for $\dot{\gamma}\left(  0\right)  $ is divergent at large $N$. So
$\dot{\gamma}\left(  0\right)  +2\dot{\beta}\left(  0\right)  =-1$ is finite,
but $\dot{x}\left(  0\right)  $ diverges. Therefore from Eq.(\ref{slope}) we
find that the slope of $f\left(  x\right)  $ at $x=1/2$ is exactly zero%
\begin{equation}
\left(  \partial_{x}f\left(  x\right)  \right)  _{x=1/2}=0.
\end{equation}
So, after all the off-shell function $f\left(  x\right)  $ is continuous and
differentiable at $x=1/2.$ With this result on the slope, the expansion in
Eq.(\ref{expansion}), and the plot in Fig.2, we have basically understood the
function $f\left(  x\right)  $.

This discussion provides the most comprehensive result for the off-shell
scattering amplitude of four tachyons produced so far in string field theory.

\section{Computational techniques}

In this paper we need to evaluate the quantities $\omega,b_{o},b_{e}%
,c_{e},c_{o},M_{ee^{\prime}},M_{oo^{\prime}},\tilde{M}_{ee^{\prime}},\tilde
{M}_{oo^{\prime}},$ and various determinants, as defined in the previous
sections. These are examples of \ more general computations that come up in
MSFT which are generically of the type
\begin{align}
&  \left(  F\left(  \bar{t}t\right)  \right)  _{oo^{\prime}},\;\left(
F\left(  t\bar{t}\right)  \right)  _{ee^{\prime}},\;\left(  tF\left(  \bar
{t}t\right)  \right)  _{eo},~\left(  F\left(  \bar{t}t\right)  \kappa
^{-1/2}v\right)  _{o},\;\left(  tF\left(  \bar{t}t\right)  \kappa
^{-1/2}v\right)  _{e},\;\\
&  \det\left(  F\left(  \bar{t}t\right)  \right)  ,~\bar{v}\kappa_{o}%
^{-1/2}F\left(  \bar{t}t\right)  \kappa_{o}^{-1/2}v,\;\bar{v}\kappa_{o}%
^{-1/2}F\left(  \bar{t}t\right)  \kappa_{o}F\left(  \bar{t}t\right)
\kappa_{o}^{-1/2}v,~etc.
\end{align}
where $F\left(  z\right)  $ can be any function of the matrix $\bar{t}t$ or
$t\bar{t}.$ In the present paper the functions $F$ of interest are
$M=\frac{1-\bar{t}t}{3+\bar{t}t}$ and $\tilde{M}=\frac{1-\bar{t}t}{1+3\bar
{t}t}$ and other simple ones constructed from them, such as $\frac{4}%
{3+\bar{t}t}=1-M,$ $\left(  1-M^{2}\right)  ,$ etc. As seen in the full $\tau
$-dependent $\alpha\left(  \tau\right)  ,\beta\left(  \tau\right)  $ etc.
there are more involved combinations of $F\left(  \bar{t}t\right)  $ and the
matrix $e^{-\tau\kappa}$ which can appear in explicit calculations. The
techniques discussed below do not directly apply to compute such quantities
fully, but do apply to compute the expansion coefficients analytically and
exactly when these quantities are expanded in powers of $e^{-\tau}.$ This is
how we made progress in this paper.

In our MSFT approach such quantities are initially defined in terms of the
regulated vector $v_{o}$ and the regulated matrix $t_{eo}$. By inserting the
values of $v_{o},t_{eo}$ for any number of modes $2N,$ all of these quantities
take on explicit numerical values. Generally it is seen that a few modes
already give an answer pretty close to the $N\rightarrow\infty$ limit,
therefore for a quick estimate of non-divergent quantities it is generally
sufficient to use finite matrices for just a few modes $N$. With this approach
we can evaluate any quantity approximately fairly easily.

The large $N$ limit is sometimes subtle because of anomalies. Generally
anomalies arise in the form of a zero multiplied by an infinity that comes
from an infinite sum over matrix indices (the modes) producing $0\times\infty
$, so that there is a subtle finite contribution. If one is not careful the
zero is first evaluated and the infinite sum that is later performed still
gives zero, thus missing the finite contribution. To avoid anomaly subtleties
in the $N\rightarrow\infty$ limit we use the regulated version of
$v_{o},t_{eo}$ in all computations (see \cite{BM1}\cite{BKM3}), and take the
$N\rightarrow\infty$ only at the end of the computation. The regulated
matrices $v_{o},t_{eo}$ satisfy certain algebraic relations, also shared by
the unregulated matrices, that make it possible to perform certain
computations analytically. It has been demonstrated that this is a correct
regulator consistent with other computational techniques in string theory.

In certain computations, or parts of it, there are no anomalies, and it
becomes possible to work with the infinite unregulated matrices $v_{e},t_{eo}
$ directly at $N=\infty$. Then it is possible to make a transformation to the
basis that diagonalizes the infinite unregulated matrix $t_{eo}$ and do the
computations in the $N=\infty$ diagonal basis. At $N=\infty$ the eigenvalues
of $t_{eo}$ are continuous and are parameterized by a continuous variable
$\kappa,$ and furthermore the transformation is known explicitly \cite{DLMZ}.
As we will discuss in detail in the next section, at $N=\infty$ the discrete
and continuous bases are just different bases of SL$\left(  2,R\right)  $ in
the $j\left(  j+1\right)  =0$ unitary representation, in which $t_{eo}$ can be
understood as the matrix elements of the operator $t=\tanh\left(  \frac{\pi
}{4}\left(  L_{1}+L_{-1}\right)  \right)  $ in the discrete basis that
diagonalizes $L_{0}$. The continuous basis is the basis in which this operator
is diagonal.

A priori it is hard to know when to expect an anomaly if one works directly at
$N=\infty$. Therefore the regulator provided in MSFT is indispensable to
insure that such subtleties will not spoil a computation. So one must be wary
when using the continuous basis directly at $N=\infty$. We emphasize that the
discrete Moyal basis with a regulator is the safe way to proceed in general,
but one can make a transformation to the continuous basis for specific
computations to evaluate infinite sums in parts of computations. This is the
sense in which we use the continuous basis in the next subsection, and indeed
we will find good use for it in some of the following computations.

\subsection{Computations via transformation to continuous basis}

In the following computations we are directly at $N=\infty$ and work with the
unregulated fundamental matrices $T,R,v,w$ in MSFT that encode string joining
(for a particular regularization see footnote \ref{regulator})
\begin{align}
w_{e} &  =\sqrt{2}\left(  i\right)  ^{-e+2},\;v_{o}=\frac{2\sqrt{2}}{\pi}%
\frac{\left(  i\right)  ^{o-1}}{o},\;\ t_{eo}=\sqrt{e}T_{eo}\frac{1}{\sqrt{o}%
},\label{v}\\
T_{eo} &  =\frac{4o\left(  i\right)  ^{o-e+1}}{\pi\left(  e^{2}-o^{2}\right)
},\;R_{oe}=\frac{4e^{2}\left(  i\right)  ^{o-e+1}}{\pi o\left(  e^{2}%
-o^{2}\right)  },\;\label{TR}%
\end{align}
As explained in \cite{BM2}\cite{B2}, the matrix $t$ is diagonalized
$t_{eo}=\left(  V_{e}\tau\bar{V}_{o}\right)  _{eo}$ by the orthogonal matrices
$V_{o},V_{e}$ that act on the odd/even sides (a bar means transpose). The
eigenspace is labelled by $\kappa$ with eigenvalues\footnote{This continuous
parameter $\kappa$ should not be confused with the diagonal matrix $\kappa$ of
Eq.(\ref{kappa}). As much as possible we are trying to keep the notations that
were introduced in different papers by independent authors. We hope the reader
will discern them in the proper contexts.} $\tau_{\kappa}$. The label $\kappa$
is continuous (can happen only for infinite matrices); it is\ in the range
$0\leq\kappa\leq\infty,$ and the eigenvalues are given by \cite{RZ}\cite{DLMZ}
$\tau_{\kappa}=\tanh\left(  \pi\kappa/4\right)  .$ The orthogonal matrices
have matrix elements which are functions of $\kappa$ and the index $o,e,$
namely $\left(  V_{o}\right)  _{o\kappa}\equiv V_{o}\left(  \kappa\right)  ,$
and $\left(  V_{e}\right)  _{e\kappa}=V_{e}\left(  \kappa\right)  .$ Thus, the
matrix equation $t_{eo}=\left(  V_{e}\right)  _{e\kappa}\tau_{\kappa}\left(
\bar{V}_{o}\right)  _{\kappa o}$ is written as%
\begin{equation}
t_{eo}=\int_{0}^{\infty}d\kappa V_{e}\left(  \kappa\right)  \tanh\left(
\pi\kappa/4\right)  V_{o}\left(  \kappa\right)
\end{equation}
The functions $V_{e}\left(  \kappa\right)  ,V_{o}\left(  \kappa\right)  $
satisfy orthogonality relations that correspond to the orthogonality
conditions on the matrices $V_{e},V_{o}.$ These functions are given explicitly
by the generating functions in Eq.(\ref{vevo}). In the next section we will
clarify the role of these functions as the overlaps $<e|\kappa>=V_{e}\left(
\kappa\right)  /\sqrt{2}$ and $<o|\kappa>=V_{o}\left(  \kappa\right)
/\sqrt{2}$ between the states of two different bases of the $j=0$
representation of SL$\left(  2,R\right)  $ generated by the Virasoro
generators $L_{0},L_{\pm1}.$ The first basis is the familiar one $|n>$
labelled by the even or odd eigenvalues $n=\left(  e,o\right)  $ of $L_{0}.$
The second basis $|\kappa>$ diagonalizes $\left(  L_{+1}+L_{-1}\right)
|\kappa>=\kappa|\kappa>.$

By evaluating the generating functions in Eq.(\ref{vevo}) at $\kappa=0$ and
expanding in powers of $z,$ one can see that $V_{e}\left(  0\right)  =0$,
while $V_{o}\left(  0\right)  $ is finite and related to $v_{o}$ as
$V_{o}\left(  0\right)  =\frac{1}{2}\sqrt{\pi o}v_{o}$. Thus $v_{o}$ is
directly related to the state $|\kappa=0>$. We have evaluated the non-trivial
infinite sum $\sum_{o}\frac{V_{o}\left(  \kappa\right)  }{\sqrt{o}}v_{o}$ in
the next section. We give it below along with a few of the $V_{e}\left(
\kappa\right)  ,V_{o}\left(  \kappa\right)  $ that are needed in our
computations
\begin{align}
V_{1}\left(  \kappa\right)   &  =\sqrt{\frac{\kappa}{\sinh\frac{\pi\kappa}{2}%
}},\;V_{2}\left(  \kappa\right)  =\frac{\kappa^{2}}{\sqrt{2\kappa\sinh
\frac{\pi\kappa}{2}}}\;,\;V_{3}\left(  \kappa\right)  =\frac{-\kappa+\frac
{1}{2}\kappa^{3}}{\sqrt{3\kappa\sinh\frac{\pi\kappa}{2}}},\\
V_{4}\left(  \kappa\right)   &  =\frac{\frac{1}{6}\kappa^{2}\left(  \kappa
^{2}-8\right)  }{\sqrt{4\kappa\sinh\frac{1}{2}\pi\kappa}},\;V_{5}\left(
\kappa\right)  =\frac{\kappa\left(  1-\frac{5}{6}\kappa^{2}+\frac{1}{24}%
\kappa^{4}\right)  }{\sqrt{5\kappa\sinh\frac{1}{2}\pi\kappa}},\;\sum_{o}%
\frac{V_{o}\left(  \kappa\right)  }{\sqrt{o}}v_{o}=\sqrt{\frac{\tanh\frac
{\pi\kappa}{4}}{\kappa\cosh^{2}\frac{\pi\kappa}{4}}}. \label{Vv}%
\end{align}

With this preparation we are ready to compute the desired generic quantities.
Applying the diagonalization of $t_{eo}$ we have just described we can write
$t=V_{e}\tau\bar{V}_{o},~$ $\bar{t}t=V_{o}\tau^{2}\bar{V}_{o}$ and $t\bar
{t}=V_{e}\tau^{2}\bar{V}_{e}$ and express the desired quantities as integrals
\begin{align}
\left(  F\left(  \bar{t}t\right)  \right)  _{oo^{\prime}} &  =\int_{0}%
^{\infty}d\kappa V_{o}\left(  \kappa\right)  F\left(  \tanh^{2}\left(
\frac{\pi\kappa}{4}\right)  \right)  V_{o^{\prime}}\left(  \kappa\right)
\label{intf}\\
\left(  F\left(  t\bar{t}\right)  \right)  _{ee^{\prime}} &  =\int_{0}%
^{\infty}d\kappa V_{e}\left(  \kappa\right)  F\left(  \tanh^{2}\left(
\frac{\pi\kappa}{4}\right)  \right)  V_{e^{\prime}}\left(  \kappa\right)  \\
\left(  tF\left(  \bar{t}t\right)  \right)  _{eo} &  =\int_{0}^{\infty}d\kappa
V_{e}\left(  \kappa\right)  \tanh\left(  \frac{\pi\kappa}{4}\right)  F\left(
\tanh^{2}\left(  \frac{\pi\kappa}{4}\right)  \right)  V_{o}\left(
\kappa\right)  \\
\left(  F\left(  \bar{t}t\right)  \kappa^{-1/2}v\right)  _{o} &  =\int
_{0}^{\infty}d\kappa V_{o}\left(  \kappa\right)  F\left(  \tanh^{2}\left(
\frac{\pi\kappa}{4}\right)  \right)  \sqrt{\frac{\tanh\frac{\pi\kappa}{4}%
}{\kappa\cosh^{2}\frac{\pi\kappa}{4}}}\\
\left(  tF\left(  \bar{t}t\right)  \kappa^{-1/2}v\right)  _{e} &  =\int
_{0}^{\infty}d\kappa V_{e}\left(  \kappa\right)  \tanh\left(  \frac{\pi\kappa
}{4}\right)  F\left(  \tanh^{2}\left(  \frac{\pi\kappa}{4}\right)  \right)
\sqrt{\frac{\tanh\frac{\pi\kappa}{4}}{\kappa\cosh^{2}\frac{\pi\kappa}{4}}}\\
\bar{v}\kappa_{o}^{-1/2}F\left(  \bar{t}t\right)  \kappa_{o}^{-1/2}v &
=\int_{0}^{\infty}d\kappa\sqrt{\frac{\tanh\frac{\pi\kappa}{4}}{\kappa\cosh
^{2}\frac{\pi\kappa}{4}}}F\left(  \tanh^{2}\left(  \frac{\pi\kappa}{4}\right)
\right)  \sqrt{\frac{\tanh\frac{\pi\kappa}{4}}{\kappa\cosh^{2}\frac{\pi\kappa
}{4}}},\label{intvv}\\
\det F\left(  \bar{t}t\right)   &  =\exp\int_{0}^{\infty}d\kappa\rho\left(
\kappa\right)  \ln\left[  F\left(  \tanh^{2}\left(  \frac{\pi\kappa}%
{4}\right)  \right)  \right]  \label{detf}%
\end{align}
In the infinite $N$ limit we have $\det\left(  F\left(  t\bar{t}\right)
\right)  =\det\left(  F\left(  \bar{t}t\right)  \right)  .$ The quantity
\begin{equation}
\rho\left(  \kappa\right)  =\left(  \frac{1}{2\pi}\ln\left(  2Ne^{\Delta
}\right)  -\frac{1}{4\pi}\left(  \psi\left(  \frac{i\kappa}{2}\right)
+\psi\left(  -\frac{i\kappa}{2}\right)  \right)  \right)
\end{equation}
which appears in the evaluation of determinants will be derived in the next
section in Eq.(\ref{rho}). Here $\psi\left(  z\right)  =\partial_{z}\ln
\Gamma\left(  z\right)  $ is the logarithmic derivative of the gamma function,
and $\Delta$ depends on the regulator. It could be chosen as zero if we
compare to a particular regulator in the discrete basis of MSFT. The same
result has been obtained in a very different type of calculation in
\cite{BK}\cite{FKM}. The term that contains the number of modes $2N$ is the
leading term independent of $\kappa.$ In physical computations, as $N$ goes to
infinity this factor cancels magically among matter and ghost determinants as
long as $d=26,$ and the bare coupling $g_{0}$ is rewritten in terms of the
on-shell tachyon coupling $g.$ Hence the non-leading part in terms of the
function $\psi\left(  z\right)  $ is the crucial part that contributes in
physical processes.

A more complicate type of quantity that we needed in Eq.(\ref{gammadot}) takes
the form of a double integral
\begin{equation}
\bar{v}\kappa_{o}^{-1/2}F\left(  \bar{t}t\right)  \kappa_{o}F\left(  \bar
{t}t\right)  \kappa_{o}^{-1/2}v=\int_{0}^{\infty}d\kappa\int_{0}^{\infty
}d\kappa^{\prime}\left(
\begin{array}
[c]{c}%
\sqrt{\frac{\tanh\frac{\pi\kappa}{4}}{\kappa\cosh^{2}\frac{\pi\kappa}{4}}%
}F\left(  \tanh^{2}\left(  \frac{\pi\kappa}{4}\right)  \right)  ~\left(
\kappa_{o}\right)  _{\kappa\kappa^{\prime}}\\
F\left(  \tanh^{2}\left(  \frac{\pi\kappa^{\prime}}{4}\right)  \right)
\sqrt{\frac{\tanh\frac{\pi\kappa^{\prime}}{4}}{\kappa\cosh^{2}\frac{\pi
\kappa^{\prime}}{4}}}%
\end{array}
\right)  \label{vfkfv}%
\end{equation}
where
\begin{equation}
\left(  \kappa_{o}\right)  _{\kappa\kappa^{\prime}}=\sum_{o}V_{o}\left(
\kappa\right)  oV_{o}\left(  \kappa^{\prime}\right)  =\frac{1}{2}<\kappa
|L_{0}|\kappa^{\prime}>+\frac{1}{2}<\kappa|L_{0}|-\kappa^{\prime
}>.\label{VkoV}%
\end{equation}
The form of $\left(  \kappa_{o}\right)  _{\kappa\kappa^{\prime}}$ in terms of
matrix elements of $L_{0}$ in the $\kappa$ basis is shown in the next section.
We apply this formula to compute $\dot{\gamma}\left(  0\right)  $ in
Eq.(\ref{gammadot}). The integrations in Eq.(\ref{vfkfv}) are well behaved for
$F\left(  \bar{t}t\right)  =\left(  1+\bar{t}t\right)  ^{-1}$. But the
infinite sum $\left(  \kappa_{o}\right)  _{\kappa\kappa^{\prime}}$ is shown to
diverge in Eqs.(\ref{klok}-\ref{klk}). It is evaluated there more carefully by
inserting a regulator in the form $\left(  e^{-\tau\kappa_{o}}\kappa
_{o}\right)  _{\kappa\kappa^{\prime}}.$ It is finite at finite $\tau,$ but
when $\tau$ is small it behaves like $1/\tau$ times a quickly oscillating
factor. The quick oscillations are not sufficient to overcome the divergence
of the $1/\tau$ factor. Therefore, this leads to the divergent result for
$\dot{\gamma}\left(  0\right)  $ in Eq.(\ref{gammadot}) which we used in the
previous section.

These integrals seem difficult, but can be easily performed with an algebraic
computer program, such as Mapple or Mathematica. For example we obtain
\begin{align}
\int_{0}^{\infty}d\kappa V_{1}\left(  \kappa\right)  \frac{2}{3+\tanh^{2}%
\frac{\pi\kappa}{4}}\sqrt{\frac{\tanh\frac{\pi\kappa}{4}}{\kappa\cosh^{2}%
\frac{\pi\kappa}{4}}}  &  =0.544\,33=\frac{2\sqrt{2}}{3\sqrt{3}}\\
\int_{0}^{\infty}d\kappa V_{2}\left(  \kappa\right)  \frac{2\tanh\frac
{\pi\kappa}{4}}{3+\tanh^{2}\frac{\pi\kappa}{4}}\sqrt{\frac{\tanh\frac
{\pi\kappa}{4}}{\kappa\cosh^{2}\frac{\pi\kappa}{4}}}  &  =0.222\,22=\frac
{2}{9}\\
\int_{0}^{\infty}d\kappa\sqrt{\frac{\tanh\frac{\pi\kappa}{4}}{\kappa\cosh
^{2}\frac{\pi\kappa}{4}}}\frac{2}{3+\tanh^{2}\frac{\pi\kappa}{4}}\sqrt
{\frac{\tanh\frac{\pi\kappa}{4}}{\kappa\cosh^{2}\frac{\pi\kappa}{4}}}  &
=0.523\,25=\ln\frac{27}{16} \label{intvv2}%
\end{align}

We now turn to the more specialized cases of relevance for our computations in
this paper. We would like to evaluate the scalar $\omega=\bar{v}\kappa
_{o}^{-1/2}\frac{2}{3+\bar{t}t}\kappa_{o}^{-1/2}v,$ the vectors $b_{o}=\left(
\frac{2}{3+\bar{t}t}\kappa_{o}^{-1/2}v\right)  _{o}$ and $b_{e}=\left(
t\frac{2}{3+\bar{t}t}\kappa_{o}^{-1/2}v\right)  _{e}$ and matrix elements
$M_{oo^{\prime}}=\left(  \frac{1-\bar{t}t}{3+\bar{t}t}\right)  _{oo^{\prime}}%
$, $M_{ee^{\prime}}=\left(  \frac{1-t\bar{t}}{3+t\bar{t}}\right)
_{ee^{\prime}}$, $\tilde{M}_{oo^{\prime}}=\left(  \frac{\bar{t}t-1}{1+3\bar
{t}t}\right)  _{oo^{\prime}}$, $\tilde{M}_{ee^{\prime}}=\left(  \frac{t\bar
{t}-1}{1+3t\bar{t}}\right)  _{ee^{\prime}}$. By applying the above integral
formulas we obtained easily the following values which are the only ones we
actually needed in this paper to perform the expansions of $\alpha\left(
\tau\right)  ,\beta\left(  \tau\right)  ,\gamma\left(  \tau\right)
,a_{4}\left(  \tau\right)  $ in powers of $e^{-\tau}$ up to order $O\left(
e^{-6\tau}\right)  $
\begin{align}
b_{1}  &  =\frac{2\sqrt{2}}{3\sqrt{3}},\;b_{2}=\frac{2}{9},\;b_{3}%
=-\frac{22\sqrt{2}}{3^{5}},~b_{4}=-\frac{19\sqrt{2}}{3^{5}},\;b_{5}%
=\frac{2^{3/2}67}{\left(  27\right)  ^{2}\sqrt{15}}\label{b1}\\
M_{11}  &  =\frac{5}{27},\;M_{22}=\frac{13}{3^{5}},\;M_{13}=-\frac{2^{5}%
\sqrt{3}}{\left(  27\right)  ^{2}},\;\omega=\ln\frac{27}{16},\label{M11}\\
\tilde{M}_{11}  &  =\frac{11}{27},\;\tilde{M}_{22}=\frac{19}{9\left(
27\right)  },\;\tilde{M}_{13}=-\frac{2^{4}5\sqrt{3}}{\left(  27\right)  ^{2}}.
\label{Mtilde11}%
\end{align}
We also show how to compute some determinants by using Eq.(\ref{detf}). We
especially make a point to separate the contribution associated with the
regulator $N$ which will be sent to infinity later. The steps of computation
are shown for the following explicit example
\begin{align}
\det\left(  \left(  t\bar{t}\right)  \right)   &  =e^{\frac{1}{2\pi}\ln\left(
2Ne^{\Delta}\right)  \int_{0}^{\infty}\ln\left(  \tanh^{2}\frac{\pi\kappa}%
{4}\right)  d\kappa}\times e^{\int_{0}^{\infty}\left(  -\frac{1}{4\pi}\left(
\psi\left(  \frac{i\kappa}{2}\right)  +\psi\left(  -\frac{i\kappa}{2}\right)
\right)  \right)  \ln\left(  \tanh^{2}\frac{\pi\kappa}{4}\right)  d\kappa
}\label{dets}\\
&  =\left(  2Ne^{\Delta}\right)  ^{-\frac{1}{2}}\left(  0.797\,88\right)
=\left(  2Ne^{\Delta}\right)  ^{-\frac{1}{2}}\sqrt{\frac{2}{\pi}}=\left(  \pi
Ne^{\Delta}\right)  ^{-1/2}%
\end{align}
As seen, the finite factor that comes from the non-leading term involving
$\psi$ is non-trivial. We will show later in Eq.(\ref{ttdiscrt}) how both the
leading and non-leading terms agree with the regulated discrete basis. In a
similar way we obtain%
\begin{align}
\det\left(  \frac{3+\left(  t\bar{t}\right)  }{4}\right)   &  =\left(
2Ne^{\Delta}\right)  ^{-\frac{1}{18}}\left(  0.980\,52\right) \\
\det\left(  \frac{1+3\left(  t\bar{t}\right)  }{4}\right)   &  =\left(
2Ne^{\Delta}\right)  ^{-\frac{2}{9}}\left(  0.917\,46\right) \\
\det\left(  1-\left(  {\frac{1-t\bar{t}}{3+t\bar{t}}}\,\right)  ^{2}\right)
&  =\left(  2Ne^{\Delta}\right)  ^{-\frac{1}{72}}\left(  0.993\,40\right) \\
\det\left(  1-\left(  {\frac{1-\bar{t}t}{1+3\bar{t}t}}\,\right)  ^{2}\right)
&  =\left(  2Ne^{\Delta}\right)  ^{-\frac{13}{72}}\left(  0.905\,34\right)
\end{align}
Note the magical number 26 that popped up as $13=26/2$ in the last ghost
determinant. Inserting these determinants in the expression for $a_{4}$ in
Eq.(\ref{a4}) we see that the $N$ dependence drops out at $\tau=0$, only for
$d=26,$
\begin{align}
a_{4}\left(  0 \right)   & =g^{2}\left(  \frac{27}{16}\right)  ^{3}%
{\frac{\left(  \left(  2Ne^{\Delta}\right)  ^{-\frac{13}{72}}\left(
0.905\,34\right)  \right)  ^{2}}{\left[  \left(  2Ne^{\Delta}\right)
^{-\frac{1}{72}}\left(  0.993\,40\right)  \right]  ^{26}}}\\
& =g^{2}\left(  \frac{27}{16}\right)  ^{3}\left(  0.973\,63\right)  .
\end{align}
So the result is independent of the cutoff $N,\Delta$. This is an example of
the magical role of $d=26.$ Note that $a_{4},$ as given initially in
\cite{BKM1}\cite{BKM3} has factors of additional determinants that are
actually divergent, but those get absorbed into the definition of $g$ and
produce the overall finite factor $g^{2}\left(  27/16\right)  ^{3}/4$
(multiplied by another factor of 4 because of 4 diagrams).

In the case of the bare coupling in Eq.(\ref{g}), inserting the results above
we get%
\begin{align}
g_{0}  &  =\frac{1}{2}g\left(  \frac{27}{16}\right)  ^{3/2}\left(  \left(
2Ne^{\Delta}\right)  ^{-\frac{1}{2}}\left(  0.797\,88\right)  \right)
^{-\left(  d-6\right)  /4}\left(  \left(  2N\right)  ^{-\frac{1}{18}}\left(
0.980\,52\right)  \right)  ^{d}\left(  \left(  2N\right)  ^{-\frac{2}{9}%
}\left(  0.917\,46\right)  \right)  ^{-2}\nonumber\\
&  =gc\left(  2Ne^{\Delta}\right)  ^{\frac{3}{2}}~~\text{for~}%
d=26~,\;\text{with~}c=\frac{1}{2}\left(  \frac{27}{16}\right)  ^{3/2}\left(
2.\,\allowbreak203\,0\right)  \label{bare}%
\end{align}
We see that the bare coupling diverges as $(N)^{\frac{3}{2}}$.

In the Appendix we have included many more values for the higher modes of the
vectors $b$ and matrices $M,\tilde{M}.$ Although these were not needed for our
analytic calculation, they were used to obtain the order ten result for
$F(x)$. Also the general formulas we give there are useful to compute to
arbitrary powers of $e^{-\tau}$ if one needs to refine our work in the future,
or they could be useful in other applications.

\section{Moyal Bases as the $j=0$ Representation of SL(2,R) for $L_{0}%
,L_{\pm1}$}

In this section we will study certain properties of the $j=0$ representation
of SL$\left(  2,R\right)  $ in relation to the continuous and discrete Moyal
bases. The connection of SL$\left(  2,R\right)  $ generated by $L_{0},L_{\pm
1}$ and the continuous basis is known through the work in \cite{RZ}%
\cite{DLMZ}. The relevance of the $j=0$ representation is emphasized in
\cite{belov}. Here we will study the $j=0$ representation with a different
technique, mainly by focusing on the exponentiated group element $e^{-\tau
L_{0}}$. We will extract some of the relevant properties of the $j=0$
representation that are needed in this paper. There are overlaps between our
results and those in related papers \cite{DLMZ}\cite{belov}\cite{BK}%
\cite{FKM}\cite{belov2}. We will see that all the computations in the previous
sections, and similarly more general computations, amount to various matrix
elements of functions of $L_{0},L_{\pm1}$ in the special $j=0$ representation
of SL$\left(  2,R\right)  .$

Let us first recall why we must focus on the $j=0$ representation. The
Virasoro operators $L_{0},L_{\pm1}$ are the generators of the SL$\left(
2,R\right)  $ transformations on the open string basis $X\left(  \tau
,\sigma\right)  .$ It is well known that this representation (in terms of
differential operators) forms the zero Casimir representation of SL$\left(
2,R\right)  .$ Thus the oscillator basis labelled by the integers
$n=1,2,3,\cdots$ is simply the case of the discrete series of SL$\left(
2,R\right)  $ labelled as $|j,m>,$ $m=j+1,$ $j+2,$ $j+3,$ $\cdots,$ for $j=0.
$

We will use the Hermitian combinations $Q_{1}=\frac{1}{2}\left(  L_{1}%
+L_{-1}\right)  $ and $Q_{2}=\frac{i}{2}\left(  L_{1}-L_{-1}\right)  $ that
form the SL$\left(  2,R\right)  $ Lie algebra as follows $\left[  Q_{1}%
,Q_{2}\right]  =-iL_{0},$ $\left[  L_{0},Q_{1}\right]  =iQ_{2},$ $\left[
L_{0},Q_{2}\right]  =-iQ_{1}.$

Thus, the odd/even discrete bases in MSFT correspond to the eigenstates of the
Virasoro operator $L_{0}|n>=n|n>,$ $n\geq1,$ with $n=o,e.$ The continuous
basis is given by the eigenstate of the Virasoro operator $Q_{1}=\frac{1}%
{2}\left(  L_{1}+L_{-1}\right)  $ satisfying $Q_{1}|\kappa>=\frac{\kappa}%
{2}|\kappa>.$ The change of basis is given by the functions $<o|\kappa
>\equiv\frac{1}{\sqrt{2}}V_{o}\left(  \kappa\right)  $ and $<e|\kappa
>=\frac{1}{\sqrt{2}}V_{e}\left(  \kappa\right)  .$ The eigenvalue of $Q_{1}$
can have any sign, but in MSFT it is possible to make transformations to the
range $\kappa\geq0$ thanks to the symmetry property $V_{o}\left(
-\kappa\right)  =V_{o}\left(  \kappa\right)  $ and $V_{e}\left(
-\kappa\right)  =-V_{e}\left(  \kappa\right)  $ of these functions. We
consider these as orthogonal matrices $V_{o},V_{e}$ with matrix elements
$\left(  V_{o}\right)  _{o\kappa}\equiv V_{o}\left(  \kappa\right)  $ and
$\left(  V_{e}\right)  _{e\kappa}\equiv V_{e}\left(  \kappa\right)  ,$ for
$\kappa\geq0,$ which satisfy orthogonality relations
\begin{equation}
\left(  V_{o}\bar{V}_{o}\right)  _{o_{1}o_{2}}=\delta_{o_{1}o_{2}},\;\left(
V_{e}\bar{V}_{e}\right)  _{e_{1}e_{2}}=\delta_{e_{1}e_{2}},\;\left(  \bar
{V}_{o}V_{o}\right)  _{\kappa\kappa^{\prime}}=\delta^{\left(  o\right)
}\left(  \kappa-\kappa^{\prime}\right)  ,\;\left(  \bar{V}_{e}V_{e}\right)
_{\kappa\kappa^{\prime}}=\delta^{\left(  e\right)  }\left(  \kappa
-\kappa^{\prime}\right)  .
\end{equation}
where a bar on a matrix means the transpose of the matrix. Here $\delta
^{\left(  o,e\right)  }\left(  \kappa-\kappa^{\prime}\right)  $ is basically
the usual Dirac delta function, except for some extra care when $\kappa
,\kappa^{\prime}$ are \textit{both} close to 0 (see below), while their sum
yields the usual delta function $\delta^{\left(  o\right)  }\left(
\kappa-\kappa^{\prime}\right)  +\delta^{\left(  e\right)  }\left(
\kappa-\kappa^{\prime}\right)  =2\delta\left(  \kappa-\kappa^{\prime}\right)
.$ These are equivalent to the orthogonality and completeness relations of the
discrete and continuous bases%
\begin{equation}
<n|n^{\prime}>=\delta_{nn^{\prime}},\;<\kappa|\kappa^{\prime}>=\delta\left(
\kappa-\kappa^{\prime}\right)  ,\;\sum_{n=1}^{\infty}|n><n|=1,\;\int_{-\infty
}^{\infty}d\kappa|\kappa><\kappa|=1.
\end{equation}
With the normalization given above the functions $V_{o}\left(  \kappa\right)
,V\,_{e}\left(  \kappa\right)  $ are identical to the functions $\sqrt{2}%
v_{o}\left(  \kappa\right)  ,\sqrt{2}v_{e}\left(  \kappa\right)  $ in
\cite{RZ}\cite{DLMZ}, including an overall normalization of $\sqrt{2}$
(consistent with counterpart orthogonal transformations in the discrete basis
as explained in \cite{BM2}). They are given by the following generating
functions (in the range $\kappa\geq0$)
\begin{equation}
\sum_{o}\frac{V_{o}\left(  \kappa\right)  \left(  \tan z\right)  ^{o}}%
{\sqrt{o}}=\frac{\sinh\left(  \kappa z\right)  }{\sqrt{\kappa\sinh\left(
\frac{\pi\kappa}{2}\right)  }},\;\;\sum_{e}\frac{V_{e}\left(  \kappa\right)
\left(  \tan z\right)  ^{e}}{\sqrt{e}}=\frac{\cosh\left(  \kappa z\right)
-1}{\sqrt{\kappa\sinh\left(  \frac{\pi\kappa}{2}\right)  }} \label{vevo}%
\end{equation}
To define the functions for negative values of $\kappa$ the right hand side of
these equations should be multiplied by the sign function $\varepsilon\left(
\kappa\right)  $ for consistency with the symmetry properties of $V_{o}\left(
-\kappa\right)  =V_{o}\left(  \kappa\right)  $ and $V_{e}\left(
-\kappa\right)  =-V_{o}\left(  \kappa\right)  .$

\bigskip A fundamental quantity in MSFT is the matrix $T_{eo}$ \cite{B1}. In
computations $T$ usually appears in the form $t_{eo}$ as given in
Eq.(\ref{v}). This matrix is diagonalized \cite{DLMZ} by $t_{eo}=\int
_{0}^{\infty}d\kappa V_{e}\left(  \kappa\right)  \tau\left(  \kappa\right)
V_{o}\left(  \kappa\right)  $ with the eigenvalues $\tau\left(  \kappa\right)
=\tanh\left(  \pi\kappa/4\right)  .$ To understand this matrix we introduce
the notion of the operator $t.$ Then $t_{eo}$ can be recognized to be just the
matrix elements of the operator $t$ in the discrete basis
\begin{equation}
t=\tanh\frac{\pi Q_{1}}{2}=\frac{1-e^{-\pi Q_{1}}}{1+e^{-\pi Q_{1}}}%
,\;t_{eo}=<e|t|o>. \label{top}%
\end{equation}
This is proven by introducing identity in the $\kappa$ basis and writing
\begin{align}
t_{eo}  &  =<e|t|o>=\int_{-\infty}^{\infty}d\kappa<e|t|\kappa><\kappa|o>\\
&  =\int_{-\infty}^{\infty}d\kappa\frac{1}{\sqrt{2}}V_{o}\left(
\kappa\right)  \tanh\frac{\pi\kappa}{4}\frac{1}{\sqrt{2}}V_{e}\left(
\kappa\right) \\
&  =\int_{0}^{\infty}d\kappa V_{o}\left(  \kappa\right)  \tanh\frac{\pi\kappa
}{4}V_{e}\left(  \kappa\right)  ,
\end{align}
where we have used the symmetry of the functions in going from the second line
to the third. By the same argument we can show $t_{ee^{\prime}}=0$ and
$t_{oo^{\prime}}=0.$ So the matrix representation of the operator $t$ is
block-off-diagonal in the even/odd basis.

Now we note the special role of the $\kappa=0$ state and its relation to
$v_{o}$. The vector $v_{o}$ is a crucial state in MSFT as seen in the
calculations in this paper. Its special role in relation to anomalies first
emerged in \cite{B1}\cite{BM1} as a state closely related to the midpoint of
the string. From the generating functions in Eq.(\ref{vevo}) it is seen that
at $\kappa=0$ the even functions vanish $V_{e}\left(  0\right)  =0$, while the
odd ones $V_{o}\left(  0\right)  $ are finite,
\begin{equation}
V_{e}\left(  0\right)  =0,\;\;V_{o}\left(  0\right)  =\frac{1}{2}\sqrt{\pi
o}v_{o}. \label{V(0)}%
\end{equation}
where $v_{o}$ is given in Eq.(\ref{v}). It is a normalized vector $\sum
_{o\geq1}\bar{v}_{o}v_{o}=1$ that is a zero mode of the matrix $T:~$
$\sum_{o\geq1}T_{eo}v_{o}=0$. Equivalently, $V_{o}\left(  0\right)  $ is the
zero mode of the matrix $t_{eo}.$ This is easily seen from the expression
Eq.(\ref{top}) for the $t$ operator acting on the zero eigenstate in the kappa
basis $Q_{1}|0>=0$
\begin{equation}
t_{eo}V_{o}\left(  0\right)  =0\;\;\longleftrightarrow\;e^{-\pi Q_{1}}|0>=|0>.
\end{equation}
Note that we can write $v_{o}$ in terms of the $\kappa=0$ state as follows%
\begin{equation}
(\kappa_{o}^{-1/2}v)_{o}=\left(  2/\sqrt{\pi}\right)  <o|\frac{1}{L_{0}}|0>.
\end{equation}

We can use the above interpretation of the operator $t$ and and its zero mode
$v_{o}$ to give an SL$\left(  2,R\right)  $ operator representation of the
computations we performed in the previous section. In particular, we see that
$b_{o},b_{e}$ of Eqs.(\ref{bobe}) can now be written as matrix elements in the
$j=0$ representation of SL$\left(  2,R\right)  $%
\begin{equation}
b_{o}=\frac{2}{\sqrt{\pi}}<o|\frac{2}{3+t^{2}}\frac{1}{L_{0}}|0>,\;b_{e}%
=\frac{2}{\sqrt{\pi}}<e|\frac{2t}{3+t^{2}}\frac{1}{L_{0}}|0>,
\end{equation}
Note that $<e|\frac{2}{3+t^{2}}\frac{1}{L_{0}}|0>=0$ since $<e|0>=0.$ Hence we
can remove the bras and write the states%
\begin{equation}
|b_{o}>=\frac{2}{\sqrt{\pi}}\frac{2}{3+t^{2}}\frac{1}{L_{0}}|0>,\;|b_{e}%
>=\frac{2}{\sqrt{\pi}}\frac{2t}{3+t^{2}}\frac{1}{L_{0}}|0>
\end{equation}
where now the $o,e$ indices may be interpreted as an odd or even number of
powers of the operator $t.$ Similarly, the quantities $\alpha\left(
\tau\right)  ,\beta\left(  \tau\right)  ,\gamma\left(  \tau\right)  $ that
appear in the 4-tachyon amplitude in Eqs.(\ref{alpha}-\ref{gamma}) become
expectation values in the $\kappa=0$ states. For example%
\begin{align}
\beta\left(  \tau\right)   &  =\bar{b}P^{\left(  o\right)  }e^{-\tau\kappa
}\left(  1-Me^{-\tau\kappa}Me^{-\tau\kappa}\right)  ^{-1}P^{\left(  o\right)
}b,\\
&  =\frac{4}{\pi}<0|\frac{1}{L_{0}}\frac{2}{3+t^{2}}e^{-\tau L_{0}}\left(
1-\frac{1-t^{2}}{3+t^{2}}e^{-\tau L_{0}}\frac{1-t^{2}}{3+t^{2}}e^{-\tau L_{0}%
}\right)  ^{-1}\frac{2}{3+t^{2}}\frac{1}{L_{0}}|0>.
\end{align}
Once we notice these forms we are tempted to use the commutation rules of the
operators of SL$\left(  2,R\right)  $ to simplify and compute these
quantities. In particular note that $e^{-\omega Q_{1}}L_{0}e^{\omega Q_{1}%
}=L_{0}\cos\omega+iQ_{2}\sin\omega$ gives the following interesting properties
of $t$ and $e^{-\pi Q_{1}}$ for $\omega=\pi$%
\begin{equation}
e^{-\pi Q_{1}}L_{0}=-L_{0}e^{-\pi Q_{1}},~tL_{0}=L_{0}t^{-1},~\frac{1}{L_{0}%
}t=t^{-1}\frac{1}{L_{0}},\;t\frac{1}{L_{0}}=\frac{1}{L_{0}}t^{-1},~etc.
\end{equation}
Using these it appears as if we may perform a number of simplifications very
efficiently. Alas, there is the problem of anomalies precisely because of the
existence of the $\kappa=0$ state on which $t$ vanishes. So its inverse cannot
be used in a cavalier fashion. In fact not all of the above formulas are
necessarily valid in general since they depend on which state they are
applied. When we introduce a complete set of intermediate states between
operators, the $\kappa=0$ state must appear, causing problems. One can easily
generate inconsistent results from the same formula with naive manipulations
of $t$. Examples of inconsistencies can be easily constructed by taking matrix
elements and using naively the formulas above on the $\kappa=0$ state. The
type of formal manipulations suggested above need to be justified at every
step or otherwise replaced by the correct result.

Thus, practical computations in string field theory require a regulator at
$\kappa=0$ because the matrix $t_{eo}$ has certain associativity anomalies
that come from the zero modes reviewed above. This is precisely the zero mode
of $T_{eo}$ and the associativity anomaly issue which are intimately related
to midpoint issues in string field theory as first understood in \cite{BM1}.

A regulated version of the matrix $t_{eo}$ with many nice mathematical
properties was introduced in \cite{BM1}\cite{BM2}\cite{BKM1}. The regulated
matrices retain the properties of the operator $t$ that are always valid (see
Eq.(2.15) in \cite{BM2}). In the regularized theory, the continuous functions
$V_{o}\left(  \kappa\right)  ,V_{e}\left(  \kappa\right)  $ can be thought of
as the large $N$ limit of $N\times N$ matrices $\left(  V_{o}\right)
_{ok},\left(  V_{e}\right)  _{ek}$ with discrete values of $k$ which label the
$N$ eigenvalues $\tau_{k}$ of the regularized $N\times N$ matrix $t_{eo}%
=V_{e}\tau\bar{V}_{o}$. The large $N$ limits of the matrix elements $\left(
V_{o}\right)  _{ok},\left(  V_{e}\right)  _{ek}$ become the functions
$V_{o}\left(  \kappa\right)  ,V\,_{e}\left(  \kappa\right)  ,$ and the
discrete eigenvalues $\tau_{k}$ become $\tau_{k}\rightarrow\tau\left(
\kappa\right)  =\tanh\left(  \pi\kappa/4\right)  .$ In the limit
$N\rightarrow\infty,$ summation over discrete $k$ is replaced by $\int
_{0}^{\infty}d\kappa$ and the discrete delta function $\delta_{kk^{\prime}}$
is replaced by the continuous delta functions $\delta^{\left(  o,e\right)
}\left(  \kappa-\kappa^{\prime}\right)  $ with non-negative $\kappa
,\kappa^{\prime}\geq0,$ but with some extra care when $\kappa,\kappa^{\prime}$
are both close to zero, as we will see below.

In this section we try to relate the large $N$ limit of the regulated theory
in the discrete basis to the regulated continuous kappa basis, and from this
extract computational methods that take advantage correctly of both the
discrete and continuous bases. For this purpose we will study several
quantities by taking advantage of the SL$\left(  2,R\right)  $ group generated
by the Virasoro operators $L_{0},L_{1},L_{-1}$

\subsection{Propagator \thinspace$<\kappa|e^{-\tau L_{0}}|\kappa^{\prime}>$}

The quantity $D_{\kappa\kappa^{\prime}}\left(  \tau\right)  =<\kappa|e^{-\tau
L_{0}}|\kappa^{\prime}>$ is a representation of a group element of SL$\left(
2,R\right)  $ with Lie generators $\left(  L_{0},L_{\pm1}=Q_{1}\mp
iQ_{2}\right)  ,$ since $|\kappa>$ is the basis that diagonalizes the
generator $Q_{1}=\frac{1}{2}\left(  L_{1}+L_{-1}\right)  ,~$with $Q_{1}%
|\kappa>=\frac{\kappa}{2}|\kappa>.$ Its integral and derivatives are
quantities of interest in various explicit computations in MSFT
\begin{equation}
<\kappa|\frac{1}{L_{0}}|\kappa^{\prime}>=\int_{0}^{\infty}d\tau D_{\kappa
\kappa^{\prime}}\left(  \tau\right)  ,\;<\kappa|L_{0}|\kappa^{\prime}>=\left[
-\partial_{\tau}D_{\kappa\kappa^{\prime}}\left(  \tau\right)  \right]
_{\tau=0}.
\end{equation}
In this section we compute $D_{\kappa\kappa^{\prime}}\left(  \tau\right)  $ as
a group element of SL$\left(  2,R\right)  $ in the special representation
$j=0. $ Of course, in the discrete basis $e^{-\tau L_{0}}$ is diagonal,
therefore we expect
\begin{align}
D_{\kappa\kappa^{\prime}}\left(  \tau\right)   &  =<\kappa|e^{-\tau L_{0}%
}\left(  \sum_{e}|e><e|~+\sum_{o}~|o><o|\right)  |\kappa^{\prime}>\\
&  =\frac{1}{2}\left(  \bar{V}_{o}e^{-\kappa_{o}\tau}V_{o}\right)
_{\kappa\kappa^{\prime}}+\frac{1}{2}\left(  \bar{V}_{e}e^{-\kappa_{e}\tau
}V_{e}\right)  _{\kappa\kappa^{\prime}}. \label{Dkk}%
\end{align}
where $\kappa_{e},\kappa_{o}$ are the diagonal matrices $\kappa_{e}%
=diag\left(  2,4,6,\cdots\right)  $ and $\kappa_{o}=diag\left(  1,3,5,\cdots
\right)  ,$ and the $V_{o},V_{e}$ are treated as matrices with matrix elements
that are the functions $\left(  V_{o}\right)  _{o\kappa}=V_{o}\left(
\kappa\right)  ,$ etc.

Following the methods in \cite{qmbook}, Chapter 7.4 (leading to Eq.(7.65)), we
can derive a differential equation for $D_{\kappa\kappa^{\prime}}\left(
\tau\right)  $ as follows. We use the fact that the eigenvalue of the
quadratic Casimir operator vanishes for the states $|\kappa>$ to write
\begin{equation}
<\kappa|e^{-\tau L_{0}}\left(  \left(  L_{0}^{2}-Q_{1}^{2}-Q_{2}^{2}\right)
\right)  |\kappa^{\prime}>=0. \label{casimir}%
\end{equation}
The first two terms can be rewritten as follows
\begin{equation}
<\kappa|e^{-\tau L_{0}}L_{0}^{2}|\kappa^{\prime}>=\partial_{\tau}^{2}%
D_{\kappa\kappa^{\prime}}\left(  \tau\right)  ,\;\;<\kappa|e^{-\tau L_{0}%
}\left(  -Q_{1}^{2}\right)  |\kappa^{\prime}>=-\left(  \frac{\kappa^{\prime}%
}{2}\right)  ^{2}D_{\kappa\kappa^{\prime}}\left(  \tau\right)  .
\end{equation}
The third term $<\kappa|e^{-\tau L_{0}}\left(  -Q_{2}^{2}\right)
|\kappa^{\prime}>$ can also be computed by noting the following property,
$e^{\tau L_{0}}Q_{1}e^{-\tau L_{0}}=Q_{1}\cosh\tau+iQ_{2}\sinh\tau,$ which can
be rewritten as
\begin{equation}
e^{-\tau L_{0}}iQ_{2}=\frac{1}{\sinh\tau}Q_{1}e^{-\tau L_{0}}-\frac{\cosh\tau
}{\sinh\tau}e^{-\tau L_{0}}Q_{1}. \label{K2relation}%
\end{equation}
Multiplying Eq.(\ref{K2relation}) with $iQ_{2}$, using $Q_{1}Q_{2}=Q_{2}%
Q_{1}-iL_{0},$ and using Eq.(\ref{K2relation}) again, gives
\begin{align}
-e^{-\tau L_{0}}Q_{2}^{2}  &  =\frac{i}{\sinh\tau}Q_{1}\left(  e^{-\tau L_{0}%
}Q_{2}\right)  -i\frac{\cosh\tau}{\sinh\tau}\left(  e^{-\tau L_{0}}%
Q_{2}\right)  Q_{1}-\frac{\cosh\tau}{\sinh\tau}e^{-\tau L_{0}}L_{0}\\
&  =\frac{1}{\sinh^{2}\tau}Q_{1}^{2}e^{-\tau L_{0}}-2\frac{\cosh\tau}%
{\sinh^{2}\tau}Q_{1}e^{-\tau L_{0}}Q_{1}+\frac{\cosh^{2}\tau}{\sinh^{2}\tau
}e^{-\tau L_{0}}Q_{1}^{2}+\frac{\cosh\tau}{\sinh\tau}\partial_{\tau}e^{-\tau
L_{0}}%
\end{align}
The matrix elements of this relation are easily evaluated in terms of
$D_{\kappa\kappa^{\prime}}\left(  \tau\right)  .$ Combining the three terms in
Eq.(\ref{casimir}) we obtain the differential equation satisfied by
$D_{\kappa\kappa^{\prime}}\left(  \tau\right)  $%
\begin{equation}
\left(  \partial_{\tau}^{2}+\frac{\cosh\tau}{\sinh\tau}\partial_{\tau}%
+\frac{\kappa^{2}+\left(  \kappa^{\prime}\right)  ^{2}-2\kappa\kappa^{\prime
}\cosh\tau}{4\sinh^{2}\tau}\right)  D_{\kappa\kappa^{\prime}}\left(
\tau\right)  =0.
\end{equation}
By taking $D_{\kappa\kappa^{\prime}}\left(  \tau\right)  =z\left(  1-z\right)
^{-i\left(  \kappa+\kappa^{\prime}\right)  /4}F\left(  z\right)  ,$ with
\begin{equation}
z=-\left(  \sinh\frac{\tau}{2}\right)  ^{-2}=-4e^{-\tau}\left(  1-e^{-\tau
}\right)  ^{-2},
\end{equation}
this differential equation becomes the hypergeometric differential equation
for $F\left(  z\right)  $
\begin{equation}
\left(  \left(  \frac{\partial}{\partial z}\right)  ^{2}+\frac{c-\left(
1+a+b\right)  z}{z\left(  1-z\right)  }\frac{\partial}{\partial z}-\frac
{ab}{z\left(  1-z\right)  }\right)  F\left(  z\right)  =0,
\end{equation}
with $F\left(  a,b;c;z\right)  \sim\operatorname{hypergeom}\left(  \left[
a,b\right]  ,\left[  c\right]  ,z\right)  ,$ and%
\begin{equation}
a=1-i\frac{\kappa}{2},\;\;b=1-i\frac{\kappa^{\prime}}{2},\;\;c=2.
\end{equation}
$\allowbreak$We need the solution that satisfies the boundary conditions that
follow from Eq.(\ref{Dkk})
\begin{equation}
D_{\kappa\kappa^{\prime}}\left(  \tau\right)  \underset{\tau\rightarrow\infty
}{\rightarrow}\frac{1}{2}V_{1}\left(  \kappa\right)  V_{1}\left(
\kappa^{\prime}\right)  e^{-\tau},\;\;D_{\kappa\kappa^{\prime}}\left(
\tau\right)  \underset{\tau\rightarrow0}{\rightarrow}\delta\left(
\kappa-\kappa^{\prime}\right)  ,
\end{equation}
where $V_{1}\left(  \kappa\right)  =\sqrt{\frac{\kappa}{\sinh\frac{\pi\kappa
}{2}}},$ as obtained from Eq.(\ref{vevo}). Therefore, the desired solution is
given by
\begin{equation}
D_{\kappa\kappa^{\prime}}\left(  \tau\right)  =\sqrt{\frac{\kappa
\kappa^{\prime}}{\sinh\frac{\pi\kappa}{2}\sinh\frac{\pi\kappa^{\prime}}{2}}%
}\frac{e^{-\tau}}{2\left(  1-e^{-\tau}\right)  ^{2}}\left(  \frac{1-e^{-\tau}%
}{1+e^{-\tau}}\right)  ^{\frac{i\kappa}{2}+\frac{i\kappa^{\prime}}{2}%
}~~F\left(  1-\frac{i\kappa}{2},1-\frac{i\kappa^{\prime}}{2};2;\frac
{-4e^{-\tau}}{\left(  1-e^{-\tau}\right)  ^{2}}\right)  . \label{D}%
\end{equation}
By using the linear identity $F\left(  a,b;c;z\right)  =\left(  1-z\right)
^{c-a-b}F\left(  c-a,c-b;c;z\right)  ,$ we see that the expression for
$D_{\kappa\kappa^{\prime}}\left(  \tau\right)  $ is real. At $\tau=\infty$ the
boundary condition is satisfied since $F\left(  a,b;c;0\right)  =1.$ For small
$\tau\sim0,$ and any $\kappa,\kappa^{\prime},$ we use the asymptotic behavior
of the hypergeometric function to write$\allowbreak\allowbreak$%
\begin{equation}
D_{\kappa\kappa^{\prime}}\left(  \tau\right)  \underset{\tau\sim0}%
{\rightarrow}\frac{1}{2^{3}}\sqrt{\frac{\kappa\kappa^{\prime}}{\sinh\frac
{\pi\kappa}{2}\sinh\frac{\pi\kappa^{\prime}}{2}}}\left[  \frac{\Gamma\left(
1+\frac{i\left(  \kappa-\kappa^{\prime}\right)  }{2}\right)  }{\Gamma\left(
1+\frac{i\kappa}{2}\right)  \Gamma\left(  1-\frac{i\kappa^{\prime}}{2}\right)
}\frac{e^{i\left(  \kappa-\kappa^{\prime}\right)  \ln\sqrt{\frac{2}{\tau}}}%
}{\frac{i}{2}\left(  \kappa-\kappa^{\prime}\right)  }~+c.c.\right]
,\allowbreak\allowbreak\label{tauzero}%
\end{equation}
where $c.c.$ is the complex conjugate. For small $\left(  \kappa^{\prime
}-\kappa\right)  \sim0$ and small $\tau\sim0$ we can write $\Gamma\left(
1+\frac{i\kappa}{2}\right)  \Gamma\left(  1-\frac{i\kappa}{2}\right)
=\frac{\pi\kappa}{2}\left(  \sinh\frac{\pi\kappa}{2}\right)  ^{-1},$ so that
the singular behavior of $D_{\kappa\kappa^{\prime}}\left(  \tau\right)  ,$
consistent with the delta function, becomes evident as follows
\begin{equation}
\left[  D_{\kappa\kappa^{\prime}}\left(  \tau\right)  \right]  _{\tau\sim
0}\simeq\frac{\sin\left(  \left(  \kappa-\kappa^{\prime}\right)  \ln
\sqrt{\frac{2}{\tau}}\right)  }{\pi\left(  \kappa-\kappa^{\prime}\right)
}=\frac{1}{2\pi}\int_{-\ln\sqrt{\frac{2}{\tau}}}^{\ln\sqrt{\frac{2}{\tau}}%
}du~e^{iu\left(  \kappa-\kappa^{\prime}\right)  }\underset{\tau\rightarrow
0}{=}\delta\left(  \kappa-\kappa^{\prime}\right)  . \label{nearzero}%
\end{equation}

We also record here the special values for $\kappa,\kappa^{\prime}=0,$ that
follow from $F\left(  a,1;2;z\right)  =\allowbreak\frac{\left(  1-z\right)
^{-a}z-\left(  1-z\right)  ^{-a}+1}{z\left(  1-a\right)  }$ at any $\tau$%
\begin{align}
&  <\kappa|e^{-\tau L_{0}}|0>=D_{\kappa0}\left(  \tau\right)  =\frac
{\sin\left[  \frac{\kappa}{2}\ln\left(  \frac{1+e^{-\tau}}{1-e^{-\tau}%
}\right)  \right]  }{\sqrt{2\pi\kappa\sinh\frac{\pi\kappa}{2}}},\;
\label{Dk0}\\
&  <0|e^{-\tau L_{0}}|0>=D_{00}\left(  \tau\right)  =\frac{1}{2\pi}\ln\left(
\frac{1+e^{-\tau}}{1-e^{-\tau}}\right)  , \label{D00}%
\end{align}
where the second line follows by setting $\kappa=0$ in the first line.

The integrals of these quantities give $<\kappa|\frac{1}{L_{0}}|0>$ and
$<0|\frac{1}{L_{0}}|0>$
\begin{equation}
<\kappa|\frac{1}{L_{0}}|0>=\int_{0}^{\infty}d\tau D_{\kappa0}\left(
\tau\right)  =\left(  \frac{\pi\tanh\frac{\pi\kappa}{4}}{16\kappa\cosh
^{2}\frac{\pi\kappa}{4}}\right)  ^{1/2},\;~<0|\frac{1}{L_{0}}|0>=\frac{\pi}%
{8}. \label{kinvlo}%
\end{equation}
This agrees with an independent calculation in Eq.(\ref{vvdiscr}).

The derivative near $\tau\sim0$ gives information about matrix elements of
$L_{0}$
\begin{align}
&  <\kappa|L_{0}|\kappa^{\prime}>=\left[  -\partial_{\tau}D_{\kappa
\kappa^{\prime}}\left(  \tau\right)  \right]  _{\tau\sim0}\\
&  =\frac{1}{2^{3}}\sqrt{\frac{\kappa\kappa^{\prime}}{\sinh\frac{\pi\kappa}%
{2}\sinh\frac{\pi\kappa^{\prime}}{2}}}\left[  \frac{\Gamma\left(
1+\frac{i\left(  \kappa-\kappa^{\prime}\right)  }{2}\right)  e^{i\left(
\kappa-\kappa^{\prime}\right)  \ln\sqrt{\frac{2}{\tau}}}}{\Gamma\left(
1+\frac{i\kappa}{2}\right)  \Gamma\left(  1-\frac{i\kappa^{\prime}}{2}\right)
\tau}+c.c.\right]  \allowbreak~ \label{klok}%
\end{align}
This diverges due to the factor $1$/$\tau.$ More simply, the quantity
$<\kappa|L_{0}|0>$ is obtained also by studying the derivative of
Eq.(\ref{Dk0}) at any $\tau$
\begin{equation}
<\kappa|e^{-\tau L_{0}}L_{0}|0>=-\partial_{\tau}D_{\kappa0}=\frac{\kappa
e^{-\tau}}{1-e^{-2\tau}}\frac{\cos\left[  \frac{\kappa}{2}\ln\left(
\frac{1+e^{-\tau}}{1-e^{-\tau}}\right)  \right]  }{\sqrt{2\pi\kappa\sinh
\frac{\pi\kappa}{2}}}.
\end{equation}
Near $\tau\sim0$ one sees that $<\kappa|L_{0}|0>$ and $<0|L_{0}|0>$ have a
singular behavior
\begin{equation}
<\kappa|L_{0}|0>\rightarrow\left(  \frac{\frac{\pi\kappa}{2}}{\sinh\frac
{\pi\kappa}{2}}\right)  ^{1/2}\frac{\cos\left[  \frac{\kappa}{2}\ln\left(
\frac{2}{\tau}\right)  \right]  }{2\pi\tau},\;<0|L_{0}|0>\rightarrow\frac
{1}{2\pi\tau}. \label{kL0}%
\end{equation}

Let us also compute $<\kappa|L_{0}|\kappa^{\prime}>$ for $\kappa^{\prime}%
\sim\kappa~$from Eq.(\ref{klok}) or from Eq.(\ref{nearzero}). We obtain the
singular behavior%
\begin{equation}
\left[  <\kappa|L_{0}|\kappa^{\prime}>\right]  _{\kappa^{\prime}\sim\kappa
}=\left[  -\partial_{\tau}D_{\kappa\kappa^{\prime}}\left(  \tau\right)
\right]  _{\tau\sim0,\kappa^{\prime}\sim\kappa}\approx\frac{\cos\left(
\left(  \kappa-\kappa^{\prime}\right)  \ln\sqrt{\frac{2}{\tau}}\right)  }%
{2\pi\tau}, \label{klkcos}%
\end{equation}
which is consistent with Eq.(\ref{kL0}), but now gives
\begin{equation}
<\kappa|L_{0}|\kappa>=\left[  -\partial_{\tau}D_{\kappa\kappa}\left(
\tau\right)  \right]  _{\tau\sim0}\approx\frac{1}{2\pi\tau} \label{klk}%
\end{equation}
at any $\kappa.$ The divergence of $<\kappa|L_{0}|\kappa^{\prime}>$ is also
discussed in \cite{DLMZ}.

The quickly oscillating factor $\exp(i\left(  \kappa-\kappa^{\prime}\right)
\ln\sqrt{\frac{2}{\tau}})$ in Eq.(\ref{klok}) suggests that only the
neighborhood of $\kappa^{\prime}\sim\kappa$ is relevant in a steepest descent
approximation when the function $<\kappa|L_{0}|\kappa^{\prime}>$ is integrated
with smooth functions $\int d\kappa\int d\kappa^{\prime}f\left(
\kappa\right)  <\kappa|L_{0}|\kappa^{\prime}>g\left(  \kappa^{\prime}\right)
$ $\sim\frac{1}{2\pi\tau}\int d\kappa f\left(  \kappa\right)  g\left(
\kappa\right)  \int dq\cos\left(  q\ln\sqrt{\frac{2}{\tau}}\right)  $, where
we have used Eq.(\ref{klkcos}), and implied a cutoff in the integration in $q$
which depends on the functions $f,g$. This expression is also divergent when
$\tau$ vanishes for well behaved behavior of $f,g$. This last form relates
directly to the quantity $\dot{\gamma}\left(  0\right)  $ in
Eq.(\ref{gammadot}) through Eqs.(\ref{vfkfv},\ref{VkoV}) as the regulator
$\tau$ is removed. We used the singular behavior $\dot{\gamma}\left(  0\right)
=\infty$ to determine the mid-range slope (i.e. differentiability) of the
off-shell factor in the 4-tachyon scattering.

\subsubsection{Computation of $\left(  \bar{V}_{e}e^{-\kappa_{e}\tau}%
V_{e}\right)  _{\kappa\kappa^{\prime}}$ and $\left(  \bar{V}_{o}e^{-\kappa
_{o}\tau}V_{o}\right)  _{\kappa\kappa^{\prime}}$}

To compute $\left(  \bar{V}_{e}e^{-\kappa_{e}\tau}V_{e}\right)  _{\kappa
\kappa^{\prime}}$ and $\left(  \bar{V}_{o}e^{-\kappa_{o}\tau}V_{o}\right)
_{\kappa\kappa^{\prime}},$ we first recall their sum as in Eq.(\ref{Dkk}), and
then extract the individual terms from the symmetry under reflections with
respect to $\kappa$ or $\kappa^{\prime}.$
\begin{align}
\left(  \bar{V}_{o}e^{-\kappa_{o}\tau}V_{o}\right)  _{\kappa\kappa^{\prime}}
&  =D_{\kappa\kappa^{\prime}}\left(  \tau\right)  +D_{\left(  -\kappa\right)
\kappa^{\prime}}\left(  \tau\right)  =D_{\kappa\kappa^{\prime}}\left(
\tau\right)  +D_{\kappa\left(  -\kappa^{\prime}\right)  }\left(  \tau\right)
,\\
\left(  \bar{V}_{e}e^{-\kappa_{e}\tau}V_{e}\right)  _{\kappa\kappa^{\prime}}
&  =D_{\kappa\kappa^{\prime}}\left(  \tau\right)  -D_{\left(  -\kappa\right)
\kappa^{\prime}}\left(  \tau\right)  =D_{\kappa\kappa^{\prime}}\left(
\tau\right)  -D_{\kappa\left(  -\kappa^{\prime}\right)  }\left(  \tau\right)
.
\end{align}
where we insert our result (\ref{D}) for $D_{\kappa\kappa^{\prime}}\left(
\tau\right)  .$

In particular, if we take $\kappa^{\prime}$=0 we get $\left(  \bar{V}%
_{e}e^{-\kappa_{e}\tau}V_{e}\right)  _{\kappa0}=0,$ while $\left(  \bar{V}%
_{o}e^{-\kappa_{o}\tau}V_{o}\right)  _{\kappa0}=2D_{\kappa0}\left(
\tau\right)  ,$ which implies
\begin{align}
\bar{v}_{o}\kappa_{o}^{1/2}e^{-\kappa_{o}\tau}V_{o}\left(  \kappa\right)   &
=\frac{2}{\sqrt{\pi}}\left(  \bar{V}_{o}e^{-\kappa_{o}\tau}V_{o}\right)
_{\kappa0}=\frac{4}{\sqrt{\pi}}<0|e^{-L_{0}\tau}|\kappa>=\frac{4}{\sqrt{\pi}%
}D_{\kappa0}\left(  \tau\right) \\
&  =\frac{2}{\pi}\frac{\sin\left[  \frac{\kappa}{2}\ln\left(  \frac
{1+e^{-\tau}}{1-e^{-\tau}}\right)  \right]  }{\sqrt{\frac{\kappa}{2}\sinh
\frac{\pi\kappa}{2}}}%
\end{align}
Taking also $\kappa=0$ gives%
\begin{align}
\bar{v}_{o}\kappa_{o}e^{-\kappa_{o}\tau}v_{o}  &  =\frac{4}{\pi}\left(
\bar{V}_{o}e^{-\kappa_{o}\tau}V_{o}\right)  _{00}=\frac{8}{\pi}<0|e^{-L_{0}%
\tau}|0>=\frac{8}{\pi}D_{00}\left(  \tau\right) \\
&  =\frac{4}{\pi^{2}}\ln\left(  \frac{1+e^{-\tau}}{1-e^{-\tau}}\right)
\end{align}

We are now ready to tackle the quantity $\bar{v}_{o}\kappa_{o}^{-1/2}%
V_{o}\left(  \kappa\right)  =\frac{2}{\sqrt{\pi}}\sum_{o>0}\bar{V}_{o}\left(
0\right)  \frac{1}{o}V_{o}\left(  \kappa\right)  $ which is needed in explicit
computations in MSFT, as in Eq.(\ref{Vv}) and the equations that follow it.
Since $V_{e}\left(  0\right)  =0,$ only the odd $V_{o}\left(  0\right)  $
contribute when the identity $1=\sum_{n}|n><n|$ is inserted in $<\kappa
|\frac{1}{L_{0}}1|0>$
\begin{equation}
\frac{1}{2}\left(  \bar{V}_{o}\frac{1}{o}V_{o}\right)  _{\kappa0}%
=<\kappa|\frac{1}{L_{0}}|0>=\left(  \frac{\pi\tanh\frac{\pi\kappa}{4}%
}{16\kappa\cosh^{2}\frac{\pi\kappa}{4}}\right)  ^{1/2},
\end{equation}
where we used the result in Eq.(\ref{kinvlo}). From this we obtain
immediately
\begin{equation}
\bar{v}_{o}\kappa_{o}^{-1/2}V_{o}\left(  \kappa\right)  =\frac{4}{\sqrt{\pi}%
}\left(  \frac{1}{2}\bar{V}_{o}\frac{1}{o}V_{o}\right)  _{\kappa0}=\left(
\frac{\tanh\frac{\pi\kappa}{4}}{\kappa\cosh^{2}\frac{\pi\kappa}{4}}\right)
^{1/2}.
\end{equation}
This result is used to relate regulated discrete basis computations of the
type $\bar{v}_{o}\kappa_{o}^{-1/2}\left(  f\left(  \bar{t}t\right)  \right)
_{oo^{\prime}}\kappa_{o^{\prime}}^{-1/2}v_{o^{\prime}}$ in the large $N$ limit
to kappa basis computations as shown in the previous section.

\subsubsection{Computation of $\rho\left(  \kappa\right)  =<\kappa|\kappa>$}

We have argued in Eq.(\ref{nearzero}) that near $\tau=0$ and $\kappa^{\prime
}\sim\kappa~$we obtain$~D_{\kappa\kappa^{\prime}}\left(  0\right)
=<\kappa|\kappa^{\prime}>=\delta\left(  \kappa-\kappa^{\prime}\right)  .$ The
quantity $\rho\left(  \kappa\right)  =<\kappa|\kappa>$ appears in the
evaluation of determinants as in
\begin{equation}
\det\left(  f\left(  t\bar{t}\right)  \right)  =\exp\left[  \int_{0}^{\infty
}d\kappa\rho\left(  \kappa\right)  \ln f\left(  \tanh^{2}\frac{\pi\kappa}%
{4}\right)  \right]  .
\end{equation}
Since we expect a singular behavior, $\rho\left(  \kappa\right)  \sim
\delta\left(  0\right)  ,$ we would like to perform such computations with a
finite number of modes $2N$ and send $N$ to infinity at the end of the
computation. Thus, on the left side we have a finite determinant that depends
on $N,$ which in principle can be computed in the regulated basis, at least
numerically. To perform such careful computations analytically by taking
advantage of the $\kappa$ basis, we define $\rho\left(  \kappa\right)
=<\kappa|\kappa>$ carefully at $\kappa^{\prime}=\kappa$ for small but finite
$\tau,$ and compare to a similar calculation with the cutoff in terms of the
number of modes $2N.$ Thus, for $\kappa^{\prime}=\pm\kappa\allowbreak$ the
cases $D_{\kappa\kappa}\left(  \tau\right)  ,D_{\kappa\left(  -\kappa\right)
}\left(  \tau\right)  ,$ as $\tau\rightarrow0,$ are used to compute the
quantity $\rho\left(  \kappa,\tau\right)  \simeq<\kappa|\kappa>$ as follows.
From Eq.(\ref{tauzero}) and from the asymptotic behavior of the hypergeometric
function at large $\left(  -z\right)  $
\begin{equation}
F\left(  a,a,c;z\right)  \rightarrow\frac{\Gamma\left(  c\right)  \left(
-z\right)  ^{-a}}{\Gamma\left(  a\right)  \Gamma\left(  c-a\right)  }\left(
\ln\left(  -z\right)  +2\operatorname{\psi}\left(  1\right)
-\operatorname{\psi}\left(  a\right)  -\operatorname{\psi}\left(  c-a\right)
\right)  ,
\end{equation}
we obtain
\begin{align}
D_{\kappa\kappa}\left(  \tau\right)  \underset{\tau\sim0}{\rightarrow}  &
=\frac{1}{4\pi}\left(  \ln\left(  \frac{2}{\tau}\right)  ^{2}%
+2\operatorname{\psi}\left(  1\right)  -\operatorname{\psi}\left(
1-\frac{i\kappa}{2}\right)  -\operatorname{\psi}\left(  1+\frac{i\kappa}%
{2}\right)  \right)  +O\left(  \tau^{2}\right)  .\\
D_{\kappa\left(  -\kappa\right)  }\left(  \tau\right)  \underset{\tau\sim
0}{\rightarrow}  &  =\frac{1}{2^{3}i\sinh\frac{\pi\kappa}{2}}\left(
\frac{\Gamma\left(  1+i\kappa\right)  }{\Gamma^{2}\left(  1+\frac{i\kappa}%
{2}\right)  }e^{i\kappa\ln\frac{2}{\tau}}~-\frac{\Gamma\left(  1-i\kappa
\right)  }{\Gamma^{2}\left(  1-\frac{i\kappa}{2}\right)  }e^{-i\kappa\ln
\frac{2}{\tau}}\right)  +O\left(  \tau^{2}\right)
\end{align}
$\allowbreak$ Note the tricky behavior near $\kappa=0:$ as long as $\kappa$ is
not close to zero, the quantity $\lim_{\tau\rightarrow0}D_{\kappa\left(
-\kappa\right)  }\left(  \tau\right)  $ oscillates rapidly and becomes
negligible as a distribution compared to $\lim_{\tau\rightarrow0}%
D_{\kappa\kappa}\left(  \tau\right)  .$ However, near $\kappa=0$ we get
\begin{equation}
D_{\left(  -\kappa\right)  \kappa}\left(  \tau\right)  \underset{\tau
\sim0,\kappa\sim0}{\rightarrow}=\frac{1}{4\pi}\ln\left(  \frac{2}{\tau
}\right)  ^{2}+\kappa^{2}\left(  -\allowbreak\frac{\pi\ln\frac{2}{\tau}}%
{24}-\frac{\ln^{3}\frac{2}{\tau}}{12\pi}+0.04\allowbreak783\right)  +O\left(
\kappa^{4}\right)  \label{Dkmk0}%
\end{equation}
Hence, if $\kappa\neq0~$not small, and $\tau$ is small$,$ we neglect
$D_{\left(  -\kappa\right)  \kappa}\left(  \tau\right)  $ and get $\left(
\bar{V}_{o}e^{-\tau o}V_{o}\right)  _{\kappa\kappa}\sim\left(  \bar{V}%
_{e}e^{-\tau e}V_{e}\right)  _{\kappa\kappa}\sim D_{\kappa\kappa}\left(
\tau\right)  \sim\rho\left(  \kappa,\tau\right)  $ with
\begin{equation}
\rho\left(  \kappa,\tau\right)  =\frac{1}{4\pi}\left(  \ln\left(  \frac
{2}{\tau}\right)  ^{2}+2\operatorname{\psi}\left(  1\right)
-\operatorname{\psi}\left(  1-\frac{i\kappa}{2}\right)  -\operatorname{\psi
}\left(  1+\frac{i\kappa}{2}\right)  \right)  . \label{rhotau}%
\end{equation}
But, close to $\kappa=0,$ because of Eq.(\ref{Dkmk0}), the behavior is
\begin{equation}
\kappa,\tau\sim0:\;\left(  \bar{V}_{o}e^{-\tau o}V_{o}\right)  _{00}=2\left[
\rho\left(  0,\tau\right)  \right]  _{\tau\sim0}=\frac{2}{4\pi}\ln\left(
\frac{2}{\tau}\right)  ^{2},~~\left(  \bar{V}_{e}e^{-\tau e}V_{e}\right)
_{00}=0. \label{vovo0}%
\end{equation}
The factor of $2$ computed here is important to establish consistency between
the regulators in the discrete and continuous bases.

We concentrate on $\rho\left(  \kappa\right)  $ for the general $\kappa.$ We
already understand its dependence on $\kappa$ explicitly, and therefore write
it in the form%
\begin{equation}
\rho\left(  \kappa\right)  =\rho\left(  0\right)  +\frac{1}{4\pi}\left(
2\operatorname{\psi}\left(  1\right)  -\operatorname{\psi}\left(
1-\frac{i\kappa}{2}\right)  -\operatorname{\psi}\left(  1+\frac{i\kappa}%
{2}\right)  \right)  .
\end{equation}
We know $\rho\left(  0\right)  $ is divergent, and is given by
\begin{equation}
\rho\left(  0\right)  =\frac{1}{2}\left(  \bar{V}_{o}V_{o}\right)
_{00}=<0|0>=\frac{\pi}{8}\left(  v_{o}\kappa_{o}v_{o}\right)  _{regulated}%
=\frac{1}{2\pi}\left(  \ln\left(  2N\right)  +\Delta^{\prime}\right)
\end{equation}
where we used a finite number of modes $2N$ as the regulator and the constant
$\Delta^{\prime}$ depends on further details of the regulator. Therefore the
full regulated $\rho\left(  \kappa\right)  $ is given by%
\begin{equation}
\rho\left(  \kappa\right)  =\frac{1}{2\pi}\left(  \ln\left(  2N\right)
+\Delta\right)  -\frac{1}{4\pi}\left(  \operatorname{\psi}\left(
1-\frac{i\kappa}{2}\right)  +\operatorname{\psi}\left(  1+\frac{i\kappa}%
{2}\right)  \right)  \label{rho}%
\end{equation}
where $\Delta$ is a constant. Note that this constant can be absorbed into a
redefinition of $N$ by writing $\ln\left(  2N\right)  +\Delta=\ln\left(
2Ne^{\Delta}\right)  $. We will further determine $\Delta=0$ by comparing to a
particular regulator in the discrete basis of MSFT in which we keep the
frequencies as simple as possible, namely $\kappa_{n}=n.$ In principle it is a
dangerous business to compare diverging quantities when using different
regulators. However, as we have already mentioned, since the $N$ dependence
will drop out in certain computations in $d=26$ then $\Delta$ will also drop
out, and hence its value may not be important.

\subsection{Exact analytic tests}

As tests of the formulas we derived above in the kappa basis we compare to
exact computations in the \textit{regulated}\footnote{\label{regulator}To
avoid anomalies we use the standard regularization of MSFT \cite{BM1}%
\cite{BM2}\cite{BKM1}. If we make the choice $\kappa_{n}=n,$ at finite $N$ the
regulated $w_{e},v_{o},T_{eo},R_{oe}$ take the form $T_{eo}=T_{eo}^{\left(
\infty\right)  }\left(  f_{e}^{N}/f_{o}^{N}\right)  ,$ $R_{oe}=R_{oe}^{\left(
\infty\right)  }\left(  f_{e}^{N}/f_{o}^{N}\right)  ,$ $w_{e}=w_{e}^{\left(
\infty\right)  }f_{e}^{N},$ $v_{o}=v_{o}^{\left(  \infty\right)  }/f_{o}^{N}$
where $T^{\left(  \infty\right)  },R^{\left(  \infty\right)  },w^{\left(
\infty\right)  },v^{\left(  \infty\right)  }$ are identical to the expressions
in the infinite limit as in Eq.(\ref{v}) but truncated as $N\times N$
matrices. The factors $f_{n}^{N}$ for $n=o,e$ are given by $f_{n}^{N}%
=\sqrt{\frac{\Gamma\left(  N+\frac{1}{2}-\frac{1}{2}n\right)  \Gamma\left(
N+\frac{1}{2}+\frac{1}{2}n\right)  }{\Gamma\left(  N+1-\frac{1}{2}n\right)
\Gamma\left(  N+1+\frac{1}{2}n\right)  }}\allowbreak\frac{\Gamma\left(
N+1\right)  }{\Gamma\left(  N+\frac{1}{2}\right)  }$. This was computed by
simply inserting $\kappa_{n}=n$ in the general regulated formulas for
arbitrary $\kappa_{n}$ given in \cite{BM1}\cite{BM2}\cite{BKM1}. Note that the
deformation factor $f_{n}^{N}$ stays pretty close to 1 (in the range
1.00-1.075) when $1\leq n\leq N,$ even for finite $N,$ but grows as $n$
approaches $2N.$ For $n=2N,$ and large $N,$ we get $f_{2N}^{N}\simeq\left(
\frac{\pi N}{2}\right)  ^{1/4}.$ The strong deformation at the edges of the
matrix can be avoided by using a different function $\kappa_{n}$ as a function
of $n.$ However, in our experience the simple choice $\kappa_{n}=n$ seems to
work well in numerical computations even at small values of $N$. In the
present example we see that it also works exactly at infinite $N$ in
comparison to the continuous basis.} discrete basis, in the limit
$N\rightarrow\infty$. The first case is $\bar{v}\kappa_{o}^{-1/2}1\kappa
_{o}^{-1/2}v$ where we compare Eq.(\ref{intvv}) versus using directly the
$v_{o}$ in \ref{v} or the regulated one in footnote \ref{regulator}. The two
ways of computing give the same answer:
\begin{align}
\left(  \bar{v}\kappa_{o}^{-1/2}1\kappa_{o}^{-1/2}v\right)  _{discrete} &
=\frac{8}{\pi^{2}}\sum_{n=1}^{\infty}\frac{1}{\left(  2n-1\right)  ^{3}%
}+O\left(  \left(  \ln N\right)  /N^{2}\right)  \underset{N\rightarrow\infty
}{\rightarrow}\frac{7\zeta\left(  3\right)  }{\pi^{2}}=0.852\,56\\
\left(  \bar{v}\kappa_{o}^{-1/2}1\kappa_{o}^{-1/2}v\right)  _{kappa} &
=\int_{0}^{\infty}d\kappa\frac{\tanh\frac{\pi\kappa}{4}}{\kappa\left(
\cosh\frac{\pi\kappa}{4}\right)  ^{2}}\times1=0.852\,56.
\end{align}
The second case is $\det\left(  \bar{t}t\right)  =\exp\left(  \int_{0}%
^{\infty}~d\kappa\rho\left(  \kappa\right)  \ln\left(  f\left(  \tanh^{2}%
\frac{\pi\kappa}{4}\right)  \right)  \right)  $ where we use $\rho\left(
\kappa\right)  $ in Eq.(\ref{dets}), while the exact computation in the
\textit{regulated }discrete basis was given in \cite{BM1}\cite{BM2} for any
$N$ as
\begin{align}
\det\left(  \bar{t}t\right)  _{discrete} &  =\left(  1+\bar{w}w\right)
_{regulated}^{-1/2}=\frac{\det\kappa_{o}}{\det\kappa_{e}}\underset{\kappa
_{n}=n}{=}\frac{1\cdot3\cdot5\cdots\left(  2N-1\right)  }{2\cdot4\cdot
6\cdots\left(  2N\right)  }\\
&  =\frac{\Gamma\left(  2N+1\right)  }{2^{2N}\Gamma^{2}\left(  N+1\right)
}\simeq\left(  \pi N\right)  ^{-1/2}\left(  1+\frac{1}{2N}\right)
^{-1/4}.\label{ttdiscrt}%
\end{align}
where the last expression is a good approximation for any $N\geq1$ (including
small $N$). Note both the leading $N$ behavior as well as the overall factor
$\left(  \pi/2\right)  ^{-1/2}\left(  2N\right)  ^{-1/2},$ both of which are
significant for the comparison of regulators in different bases. We compare
this to the computation of $\det\left(  \bar{t}t\right)  _{kappa}$ in the
kappa basis using $\rho\left(  \kappa\right)  $ as given in Eq.(\ref{dets}).
We see that the leading term $\frac{1}{2\pi}\ln\left(  2Ne^{\Delta}\right)  $
in $\rho\left(  \kappa\right)  $ reproduces the correct $N$ dependence, while
the second kappa dependent term produces precisely a factor of $\sqrt{\frac
{2}{\pi}},$ leading to the total result $\det\left(  \bar{t}t\right)
_{kappa}=\left(  \pi Ne^{\Delta}\right)  ^{-1/2}.$ Comparing to the regulated
discrete basis result in Eq.(\ref{ttdiscrt}) we learn that we need $\Delta=0$
to get agreement between the two regulated results. Then, we seem to have an
agreement between the computational procedures in the regulated discrete basis
and the regulated kappa basis. However, we must warn the reader that
$\Delta=0$ is still tentative because it is possible to change the regulator
in the discrete basis, and we have not understood yet the principle that could
fix it in either the discrete or the continuous basis. It is however
significant that we have seen examples where $N$ as well as $\Delta$ cancel
together in finite quantities at the critical dimension $d=26.$ So, perhaps
the value of $\Delta$ is not crucial as long as one is consistent in using the
same regulator everywhere.

The third case is the exact computation of $\bar{v}v$ in the regulated
discrete basis for any $N$ \cite{BM1}\cite{BM2}
\begin{align}
\left(  \bar{v}v\right)  _{discrete}  &  =\left(  \frac{\bar{w}w}{1+\bar{w}%
w}\right)  _{regulated}=1-\frac{\det\kappa_{o}}{\det\kappa_{e}}\\
&  \underset{\kappa_{n}=n}{=}1-\frac{\Gamma\left(  2N+1\right)  }{2^{2N}%
\Gamma^{2}\left(  N+1\right)  }\underset{N\rightarrow\infty}{\rightarrow}1.
\label{vvdiscr}%
\end{align}
At $N=\infty$ in the kappa basis we use $V_{o}\left(  0\right)  =\frac{1}%
{2}\sqrt{\pi o}v_{o}$ and obtain agreement with the discrete basis as follows
\begin{align}
\left(  \bar{v}v\right)  _{kappa}  &  =\frac{8}{\pi}\lim_{\kappa\rightarrow
0}\left(  \frac{1}{2}\bar{V}_{o}\frac{1}{o}V_{o}\right)  _{\kappa\kappa}%
=\frac{8}{\pi}<0|\frac{1}{L_{0}}|0>=1
\end{align}
where we used Eq.(\ref{kinvlo}).

In general we have noticed no problem in agreeing between the two bases for
finite quantities where the effects of $N$ disappear. For such quantities it
turns out that even a few modes in the discrete basis ($N\sim5-$10) gives very
reliable numerical estimates of the exact $N=\infty$ values. However we advise
care when the quantity blows up or vanishes as a power $N$ (such as
determinants). Finite $N$ estimates are not predictably good for such
quantities and furthermore they seem to be cutoff dependent (such as $\Delta
$). When divergent quantities occur in combinations in which $N$ cancels, the
various computational approaches seem to become reliable again, and
furthermore in our experience, the result seems to be well estimated again at
relatively small values of $N$.

\section{Nonperturbative Landscape}

The perturbative computations above have implications for nonperturbative
computations in MSFT such as the ones in \cite{BKM2}. There a proposal was
made for computing the true vacuum of Witten's string field theory at the
classical level. This involved solving the equation of motion that follows
from the action in Eq.(\ref{action})%
\begin{equation}
(L_{0}-1)A+{\alpha^{\prime}}g_{0}A\star A=0\,\,, \label{eom}%
\end{equation}
for fields that are independent of the midpoint (hence the D$_{25}$ brane
vacuum). This could be done by treating the energy of the midpoint (called
$\gamma$) as a perturbation and writing the rest of $L_{0}$ as a special star
product form ${\mathcal{L}_{0}}\star A+A\star{\mathcal{L}_{0}}$. All solutions
of Eq.(\ref{eom}), including the vacuum, were obtained exactly in the absence
of $\gamma$. Then it was possible to setup a perturbation series in the
midpoint energy $\gamma$ and compute analytically each term order by order.
The lowest term was obtained as an exact solution $A^{\left(  0\right)
}\left(  x_{e},p_{e}\right)  $, where $A^{\left(  0\right)  }\left(
x_{e},p_{e}\right)  $ turned out to be related, up to an energy dependent
factor $A^{\left(  0\right)  }=-\left(  2/{\alpha^{\prime}}g\right)
{\mathcal{L}_{0}}\star P,$ to the twisted butterfly projector $P=A_{b}\left(
x_{e},p_{e}\right)  $ written in the MSFT basis\footnote{A proposal based on
the butterfly that has parallels to our program appeared recently
\cite{okawa}.}. The next perturbation $A^{\left(  1\right)  }$ was also
computed explicitly, while higher orders $A^{\left(  n\right)  }$ could be
computed in a straightforward way with similar methods. Finally the energy of
the vacuum and the tension of the D$_{25}$ brane were given analytically up to
second order in $\gamma$
\begin{align}
T_{25}  &  =\frac{1}{V_{25}}S\left(  A^{\left(  0\right)  }+\varepsilon
A^{\left(  1\right)  }+\varepsilon^{2}A^{\left(  2\right)  }+\cdots\right) \\
&  =\frac{-1}{\alpha^{\prime3}g_{0}^{2}}\left(  \frac{4}{3}\nu^{3}%
+\varepsilon\frac{\nu^{2}\delta}{2}\right)  +O\left(  \varepsilon^{2}\right)
, \label{tension}%
\end{align}
with%
\begin{equation}
\nu=\frac{1}{2}-\frac{d-2}{4}\left(  \sum_{e>0}\kappa_{e}-\sum_{o>0}\kappa
_{o}\right)  ,\;\;\delta=\left(  d-2\right)  \frac{\sum_{e>0}\kappa_{e}%
w_{e}^{2}}{1+\bar{w}w}.
\end{equation}

When this result was obtained the divergent nature of $\nu$ and $\delta$ were
confusing. However the divergence of the bare coupling constant $g_{0}$ was
not noticed. We have seen in this paper in Eqs.(\ref{g},\ref{bare}) that
$g_{0}=gc\left(  2Ne^{\Delta}\right)  ^{\frac{3}{2}}.$ A quick look at
$\nu,\delta$ show that they are both linearly divergent with $\left(
2N\right)  $ as can be seen in a naive level truncation up to $\left(
2N\right)  $ by inserting $\kappa_{o}=o$, $\kappa_{e}=e$ and using the
unregulated $w_{e}=-\sqrt{2}\left(  -1\right)  ^{e/2}.$ Then the result for
$T_{25}$ is perfectly finite at each order of the midpoint energy $\gamma$
since the factors of $\left(  2N\right)  ^{3}$ cancel between numerator and
denominator in Eq.(\ref{tension}). Thus, this nonperturbative computation
should proceed by using the same regulator consistently, keeping only the
leading terms to order $\left(  2N\right)  ^{3}$ in the numerator and dropping
everything non-leading. It reminds one of the large $N$ matrix computations.
Progress on determining the vacuum state and energy using this approach will
be reported in a future paper.

The observation above resolves another puzzle as follows. We recall the
expression for $L_{0}$ that appears in the action of MSFT in Eq.(\ref{action}%
). We display the version in \cite{BKM2} including ghosts $L_{0}%
=L_{0}^{matter}+L_{0}^{ghost}\,\,,$%
\begin{align}
L_{0}^{matter}  &  =\sum_{e>0}\left(  -\alpha^{\prime}\frac{\partial^{2}%
}{\partial x_{e}^{2}}-\frac{1}{16\alpha^{\prime}}\kappa_{e}^{2}\frac
{\partial^{2}}{\partial p_{e}^{2}}+\frac{1}{4\alpha^{\prime}}\kappa_{e}%
^{2}x_{e}^{2}+4\alpha^{\prime}p_{e}^{2}\right)  +\frac{1}{2}\left(  1+\bar
{w}w\right)  \beta_{0}^{2}\nonumber\\
&  +\frac{i\sqrt{2\alpha^{\prime}}}{2}\beta_{0}\sum_{e>0}w_{e}\frac{\partial
}{\partial x_{e}}-\frac{4\alpha^{\prime}}{1+\bar{w}w}\left(  \sum_{e>0}%
w_{e}p_{e}\right)  ^{2}-\frac{d}{2}\sum_{n=1}^{2N}\kappa_{n}\,\,,\label{L0}\\
L_{0}^{ghost}  &  =i\sum_{e>0}\left(  \frac{\partial}{\partial x_{e}^{b}}%
\frac{\partial}{\partial x_{e}^{c}}+\frac{{1}}{4}\kappa_{e}^{2}\frac{\partial
}{\partial p_{e}^{b}}\frac{\partial}{\partial p_{e}^{c}}+\kappa_{e}^{2}%
x_{e}^{b}x_{e}^{c}+4p_{e}^{b}p_{e}^{c}\right) \nonumber\\
&  -\frac{i}{1+\bar{w}w}\left(  \sum_{e}w_{e}\frac{\partial}{\partial
x_{e}^{b}}\right)  \left(  \sum_{e^{\prime}}w_{e^{\prime}}\frac{\partial
}{\partial x_{e^{\prime}}^{c}}\right)  +\sum_{n=1}^{2N}\kappa_{n}\,\,.
\label{L0gh}%
\end{align}
where $\beta_{0}=-i\sqrt{2\alpha^{\prime}}\frac{\partial}{\partial\bar{x}%
}=-i\sqrt{2\alpha^{\prime}}\frac{\partial}{\partial x_{cm}}$ represents the
center of mass momentum. We had commented before in previous papers that the
terms involving $\left(  1+\bar{w}w\right)  ^{-1}$ are tricky. Even though
they appear to vanish as $N\rightarrow\infty$ they actually contribute a
finite term because infinite sums cancel the zero. The midpoint energy
$\gamma$ mentioned above is in fact just the piece of $L_{0}$ proportional to
$\left(  1+\bar{w}w\right)  ^{-1}.$

But how about the divergent term in the first line $\frac{1}{2}\left(
1+\bar{w}w\right)  \beta_{0}^{2}$ involving the center of mass momentum
$\beta_{0}?$ This divergence caused concern for some colleagues. Actually we
can explain that this is the correct behavior of this term because otherwise
there will be no center of mass momentum dependence in the large $N$ limit in
nonperturbative physics. Let us start with the cubic term $g_{0}Tr\left(
A^{3}\right)  $ and replace $A$ with $A=g_{0}^{-1}A^{\prime}$ so that it takes
the form $g_{0}^{-2}Tr\left(  A^{\prime}\right)  ^{3}.$ Doing the same to the
quadratic term, we rewrite the action with an overall $g_{0}^{-2}$ as
$S=-g_{0}^{-2}\left(  Tr\left(  \left(  1/2\alpha^{\prime}\right)  A^{\prime
}\left(  L_{0}-1\right)  A^{\prime}+\left(  1/3\right)  Tr\left(  A^{\prime
}\right)  ^{3}\right)  \right)  .$ Now it has the form of the computed
nonperturbative energy in Eq.(\ref{tension}) and, from the discussion we gave
above, we see that the numerator must behave like $N^{3}$ to compensate for
the behavior of $g_{0}=cg\left(  2Ne^{\Delta}\right)  ^{3/2}$ in the
denominator. It is clear that each $A^{\prime}$ accounts for a factor of $N$
in the numerator and also that $L_{0}$ contributes a factor of $N$ in the
quadratic term. Now, if we also include momentum dependence (or $\bar{x}$
dependence) in $L_{0}$ the only way that momentum will not be negligible is by
getting the help from the factor $\left(  1+\bar{w}w\right)  \sim2N$ in the
form that it appears in $L_{0}=\cdots+\frac{1}{2}\left(  1+\bar{w}w\right)
\beta_{0}^{2}.$ This explains that this factor has precisely the correct
behavior, and how it contributes in nonperturbative phenomena.

Of course, there is a way of eliminating the confusing factor by renormalizing
$A^{\prime}=\left(  1+\bar{w}w\right)  \tilde{A}$ and defining $\tilde{L}%
_{0}=\left(  1+\bar{w}w\right)  ^{-1}L_{0},$ and further absorbing the extra
factors in a definition $\tilde{g}_{0}=\left(  1+\bar{w}w\right)  ^{3/2}%
g_{0}.$ Then in the new $\tilde{L}_{0}$ every sum is divided by the factor
$\left(  1+\bar{w}w\right)  \sim2N$ which is reminiscent of the finite
quantities $Tr\left(  M\right)  /N$ in large $N$ matrix theories. If the
theory is redefined in this manner in terms of $\left(  \tilde{A},\tilde
{L}_{0}\right)  $ then every term of the action in every computation should be
finite at every step, just like the leading terms in large $N$ matrix theories.

\section{Conclusion}

In this paper we have demonstrated that MSFT not only agreed in great detail
with other computational approaches to Witten's string field theory, but it
also led to new results that were not obtained before. We have developed
several practical and theoretical tools on the way to the new results, and we
have indicated how certain nonperturbative computations can be conducted by
using the information provided in the present paper.

\section{Acknowledgements}

I.B. would like to thank E. Fuchs, D. Gross, I. Kishimoto, M. Kroyter, Y.
Matsuo, O. Oreshkov, S. Ramanujam, S. Samuel and W. Taylor for discussions,
correspondence or comments. I.B. is in part supported by a DOE grant
DE-FG03-84ER40168. He is grateful to the CERN TH-division and to the KITP in
Santa Barbara, for hospitality while this work was performed. I.P. is
supported in part by \ DOE grant DE-FG02-91ER-40690, and he is grateful to S.
D. Mathur for his support and encouragement.

\section{Appendix}

Anticipating future applications we compute also the following quantities by
performing the integrals, which can be done easily by using an algebraic
program. We give the results for five significant figures%
\begin{equation}
\left(  t\frac{2}{3+\bar{t}t}\kappa_{o}^{-1/2}v\right)  _{e}=b_{e}%
=10^{-2}\times\allowbreak\left(
\begin{array}
[c]{c}%
22.\,\allowbreak222\\
-11.\,\allowbreak058\\
7.\,\allowbreak198\,2\\
-5.\,\allowbreak271\,9\\
4.\,\allowbreak127\,5\\
\vdots
\end{array}
\right)  ,\;\; \label{be}%
\end{equation}%
\begin{equation}
\left(  \frac{2}{3+\bar{t}t}\kappa_{o}^{-1/2}v\right)  _{o}=b_{o}%
=10^{-2}\times\left(
\begin{array}
[c]{c}%
54.\,\allowbreak433\\
-12.\,\allowbreak804\\
6.\,\allowbreak711\,9\\
-4.\,\allowbreak418\,3\\
3.\,\allowbreak243\,2\\
\vdots
\end{array}
\right)  \label{bo}%
\end{equation}%
\begin{equation}
\left(  \bar{t}t\right)  _{oo^{\prime}}=10^{-2}\times\left(
\begin{array}
[c]{cccccc}%
40.\,\allowbreak528 & 23.\,\allowbreak399 & -14.\,\allowbreak097 &
9.\,\allowbreak871\,8 & -7.\,\allowbreak488\,1 & \cdots\\
23.\,\allowbreak399 & 85.\,\allowbreak56 & 10.\,\allowbreak464 &
-8.\,\allowbreak136\,5 & 6.\,\allowbreak611\,2 & \cdots\\
-14.\,\allowbreak097 & 10.\,\allowbreak464 & 91.\,\allowbreak684 &
6.\,\allowbreak850\,6 & -5.\,\allowbreak795 & \cdots\\
9.\,\allowbreak871\,8 & -8.\,\allowbreak136\,5 & 6.\,\allowbreak850\,6 &
94.\,\allowbreak133 & 5.\,\allowbreak106\,1 & \cdots\\
-7.\,\allowbreak488\,1 & 6.\,\allowbreak611\,2 & -5.\,\allowbreak795 &
5.\,\allowbreak106\,1 & 95.\,\allowbreak46 & \cdots\\
\vdots & \vdots & \vdots & \vdots & \vdots & \ddots
\end{array}
\right)
\end{equation}
Note the increasing diagonal, although the off-diagonals are much smaller. The
increase on the diagonal $\left(  \bar{t}t\right)  _{oo}$ quickly slows down
as $o$ increases, and stays around $160<\left(  \bar{t}t\right)  _{oo}<167$ in
the range $100<o<175$ and then makes a sharp drop reaching $\left(  \bar
{t}t\right)  _{oo}$=$2.\,\allowbreak261\,4$ at $o=185,$ and continues to drop
slowly as $o$ increases.

The following are exact computations, like the ones above, obtained directly
by using the integrals, not by inserting a truncated form of $t\bar{t}$ or
$\bar{t}t.$ If compared to what follows from the truncated $t\bar{t}$ or
$\bar{t}t$ one finds results that disagree although they are in a similar
range of values. Therefore, the contributions from the higher modes are not
always negligible.
\begin{equation}
\left(  \frac{1-t\bar{t}}{3+t\bar{t}}\right)  _{ee^{\prime}}=M_{ee^{\prime}%
}=10^{-2}\times\left(
\begin{array}
[c]{cccccc}%
5.\,\allowbreak349\,8 & -3.\,\allowbreak678\,7 & 2.\,\allowbreak753\,3 &
-2.\,\allowbreak183\,7 & 1.\,\allowbreak801\,3 & \cdots\\
-3.\,\allowbreak678\,7 & 2.\,\allowbreak893\,1 & -2.\,\allowbreak333\,7 &
1.\,\allowbreak942\,2 & -1.\,\allowbreak657\,1 & \cdots\\
2.\,\allowbreak753\,3 & -2.\,\allowbreak333\,7 & 1.\,\allowbreak973\,9 &
-1.\,\allowbreak697\,8 & 1.\,\allowbreak484\,3 & \cdots\\
-2.\,\allowbreak183\,7 & 1.\,\allowbreak942\,2 & -1.\,\allowbreak697\,8 &
1.\,\allowbreak496\,1 & -1.\,\allowbreak332\,3 & \cdots\\
1.\,\allowbreak801\,3 & -1.\,\allowbreak657\,1 & 1.\,\allowbreak484\,3 &
-1.\,\allowbreak332\,3 & 1.\,\allowbreak203\,9 & \cdots\\
\vdots & \vdots & \vdots & \vdots & \vdots & \ddots
\end{array}
\allowbreak\right)  \label{Me}%
\end{equation}%
\begin{equation}
\left(  \frac{1-\bar{t}t}{3+\bar{t}t}\right)  _{oo^{\prime}}=M_{oo^{\prime}%
}=10^{-2}\times\left(
\begin{array}
[c]{cccccc}%
18.\,\allowbreak519 & -7.\,\allowbreak603\,0 & 4.\,\allowbreak725\,9 &
-3.\,\allowbreak393\,3 & 2.\,\allowbreak628\,7 & \cdots\\
-7.\,\allowbreak603\,0 & 4.\,\allowbreak536\,9 & -3.\,\allowbreak288\,2 &
2.\,\allowbreak572\,3 & -2.\,\allowbreak106 & \cdots\\
4.\,\allowbreak725\,9 & -3.\,\allowbreak288\,2 & 2.\,\allowbreak582\,0 &
-2.\,\allowbreak124\,3 & 1.\,\allowbreak801\,2 & \cdots\\
-3.\,\allowbreak393\,3 & 2.\,\allowbreak572\,3 & -2.\,\allowbreak124\,3 &
1.\,\allowbreak808\,8 & -1.\,\allowbreak572\,5 & \cdots\\
2.\,\allowbreak628\,7 & -2.\,\allowbreak106 & 1.\,\allowbreak801\,2 &
-1.\,\allowbreak572\,5 & 1.\,\allowbreak393\,3 & \cdots\\
\vdots & \vdots & \vdots & \vdots & \vdots & \ddots
\end{array}
\right)  \label{Mo}%
\end{equation}%
\begin{equation}
\left(  \frac{t\bar{t}-1}{1+3t\bar{t}}\right)  _{ee^{\prime}}=\tilde
{M}_{ee^{\prime}}=10^{-2}\times\left(
\begin{array}
[c]{cccccc}%
7.\,\allowbreak818\,9 & -5.\,\allowbreak748\,0 & 4.\,\allowbreak536\,7 &
-3.\,\allowbreak761\,3 & 3.\,\allowbreak223\,3 & \cdots\\
-5.\,\allowbreak748\,0 & 4.\,\allowbreak657\,7 & -3.\,\allowbreak872\,5 &
3.\,\allowbreak315\,4 & -2.\,\allowbreak903\,5 & \cdots\\
4.\,\allowbreak536\,7 & -3.\,\allowbreak872\,5 & 3.\,\allowbreak326\,6 &
-2.\,\allowbreak912\,2 & 2.\,\allowbreak591\,8 & \cdots\\
-3.\,\allowbreak761\,3 & 3.\,\allowbreak315\,4 & -2.\,\allowbreak912\,2 &
2.\,\allowbreak591\,3 & -2.\,\allowbreak334\,8 & \cdots\\
3.\,\allowbreak223\,3 & -2.\,\allowbreak903\,5 & 2.\,\allowbreak591\,8 &
-2.\,\allowbreak334\,8 & 2.\,\allowbreak124\,1 & \cdots\\
\vdots & \vdots & \vdots & \vdots & \vdots & \ddots
\end{array}
\right)  \label{Mtildee}%
\end{equation}%
\begin{equation}
\left(  \frac{\bar{t}t-1}{1+3\bar{t}t}\right)  _{oo^{\prime}}=\tilde
{M}_{oo^{\prime}}=10^{-2}\times\left(
\begin{array}
[c]{cccccc}%
40.\,\allowbreak741 & 19.\,\allowbreak007 & -12.\,\allowbreak905 &
9.\,\allowbreak917\,1 & -8.\,\allowbreak119\,1 & \cdots\\
19.\,\allowbreak007 & 10.\,\allowbreak664 & 7.\,\allowbreak800\,8 &
-6.\,\allowbreak237\,6 & 5.\,\allowbreak233\,4 & \cdots\\
-12.\,\allowbreak905 & 3.\,\allowbreak288\,2 & 5.\,\allowbreak957\,3 &
4.\,\allowbreak895\,7 & -4.\,\allowbreak185\,1 & \cdots\\
9.\,\allowbreak917\,1 & -6.\,\allowbreak237\,6 & 4.\,\allowbreak895\,7 &
4.\,\allowbreak101\,4 & 3.\,\allowbreak556\,1 & \cdots\\
-8.\,\allowbreak119\,1 & 5.\,\allowbreak233\,4 & -4.\,\allowbreak185\,1 &
3.\,\allowbreak556\,1 & 3.\,\allowbreak117\,1 & \cdots\\
\vdots & \vdots & \vdots & \vdots & \vdots & \cdots
\end{array}
\right)  \label{Mtildeo}%
\end{equation}
Other matrix elements of interest are obtained from the ones above, such as
$\frac{4}{3+t\bar{t}}=1+M,\;$and$\;\frac{2\left(  1+t\bar{t}\right)  }%
{3+t\bar{t}}=1-M,$ etc.$\allowbreak$

We can also evaluate some of the quantities by using other methods and compare
to the results above. In particular, by exact summation over the infinite
modes we can express the infinite matrices $\left(  t\bar{t}\right)
_{ee^{\prime}}$ and $\left(  \bar{t}t\right)  _{oo^{\prime}}$ in terms of the
generalized hypergeometric function as follows%
\begin{align}
\left(  \bar{t}t\right)  _{oo^{\prime}}  &  =\frac{32\left(  i\right)
^{o-o^{\prime}}\sqrt{oo^{\prime}}}{\pi^{2}\left(  o^{2}-4\right)  \left(
\left(  o^{\prime}\right)  ^{2}-4\right)  }\\
&  \times\allowbreak\operatorname{hypergeom}\left(  \left[  2,1+\frac{o}%
{2},1-\frac{o^{\prime}}{2},1+\frac{o^{\prime}}{2},1-\frac{o}{2}\right]
,\left[  2+\frac{o}{2},2-\frac{o}{2},2-\frac{o^{\prime}}{2},2+\frac{o^{\prime
}}{2}\right]  ,1\right) \nonumber
\end{align}
The values of the hypergeometric function exactly agree with the results of
the integrals given above.

Also we recognize the quantities $\omega,b_{o},b_{e},M_{oo^{\prime}%
},M_{ee^{\prime}},\tilde{M}_{oo^{\prime}},\tilde{M}_{ee^{\prime}}$ from our
earlier work on the computation of the Neumann matrices by using the Moyal
product \cite{BM2}\cite{BKM3} and comparing to Neumann coefficients which were
obtained from conformal field theory \cite{GJ}. Therefore, in some sense we
already knew the result for these quantities. However, what we knew is only a
special case of the more general formulas given in Eq.(\ref{intf}%
-\ref{intvv}), and serves to confirm the general method. For example $\left(
\bar{t}t\right)  _{oo^{\prime},}\left(  t\bar{t}\right)  _{ee^{\prime}}$ or
the determinants evaluated in the text cannot be obtained by using just the
Neumann matrices. Thus, from our earlier work we extract the following results
given in terms of the vectors $A_{e},A_{o},B_{e},B_{o}$ whose numerical values
are given by the following generating functions
\begin{equation}
\left(  \frac{1+z}{1-z}\right)  ^{1/3}=1+\sum A_{e}z^{e}+i\sum A_{o}%
z^{o},\;\left(  \frac{1+iz}{1-iz}\right)  ^{2/3}=1+\sum B_{e}z^{e}+i\sum
B_{o}z^{o}.
\end{equation}
$\allowbreak$In terms of these we have the desired quantities as obtained from
\cite{BM2}\cite{BKM3} with a little algebra
\begin{align}
b_{o}  &  =\left(  \frac{2}{3+\bar{t}t}\kappa_{o}^{-1/2}v\right)  _{o}%
=\sqrt{\frac{2}{3}}\frac{A_{o}}{\sqrt{o}},\;\;\tilde{b}_{o}=\left(  \frac
{2}{1+3\bar{t}t}\kappa_{o}^{-1/2}v\right)  _{o}=\sqrt{\frac{2}{3}}\frac{B_{o}%
}{\sqrt{o}}\label{bb}\\
b_{e}  &  =\left(  t\frac{2}{3+\bar{t}t}\kappa_{o}^{-1/2}v\right)  _{e}%
=-\sqrt{2}\frac{A_{e}}{\sqrt{e}},\;\;\tilde{b}_{e}=\left(  t\frac{2}%
{1+3\bar{t}t}\kappa_{o}^{-1/2}v\right)  _{e}=-\frac{\sqrt{2}}{3}\frac{B_{e}%
}{\sqrt{e}}%
\end{align}
and%
\begin{align}
M_{oo^{\prime}}  &  =\left(  \frac{1-\bar{t}t}{3+\bar{t}t}\right)
_{oo^{\prime}}=\left\{
\begin{array}
[c]{l}%
\frac{1}{3}\left(  1-A_{o}^{2}-2\sum_{o^{\prime\prime}=1}^{o-2}A_{o^{\prime
\prime}}^{2}+2\sum_{e^{\prime\prime}=2}^{o-1}A_{e^{\prime\prime}}^{2}\right)
,~for~o=o^{\prime}~\\
\frac{1}{3}\sqrt{oo^{\prime}}\left(  \frac{\left(  A\bar{B}+B\bar{A}\right)
_{oo^{\prime}}}{o+o^{\prime}}+\frac{\left(  A\bar{B}-B\bar{A}\right)
_{oo^{\prime}}}{o-o^{\prime}}\right)  ,\;for~o\neq o^{\prime}%
\end{array}
\right. \\
M_{ee^{\prime}}  &  =\left(  \frac{1-t\bar{t}}{3+t\bar{t}}\right)
_{ee^{\prime}}=\left\{
\begin{array}
[c]{l}%
\frac{1}{3}\left(  1+A_{e}^{2}+2\sum_{e^{\prime\prime}=2}^{e-2}A_{e^{\prime
\prime}}^{2}-2\sum_{o^{\prime\prime}=1}^{e-1}A_{o^{\prime\prime}}^{2}\right)
,~for~e=e^{\prime}~\\
\frac{1}{3}\sqrt{ee^{\prime}}\left(  \frac{\left(  A\bar{B}+B\bar{A}\right)
_{ee^{\prime}}}{e+e^{\prime}}+\frac{\left(  A\bar{B}-B\bar{A}\right)
_{ee^{\prime}}}{e-e^{\prime}}\right)  ,\;for~e\neq e^{\prime}%
\end{array}
\right. \\
Y_{eo}  &  =\left(  t\frac{1}{3+t\bar{t}}\right)  _{eo}=-\frac{\sqrt{eo}%
}{4\sqrt{3}}\left(  \frac{\left(  A\bar{B}-B\bar{A}\right)  _{eo}}{e+o}%
+\frac{\left(  A\bar{B}+B\bar{A}\right)  _{eo}}{e-o}\right)
\end{align}
where a bar on a vector, such as $\bar{A},$ means its transpose. Similarly we
have
\begin{align}
\tilde{M}_{oo^{\prime}}  &  =\left(  \frac{1-\bar{t}t}{1+3\bar{t}t}\right)
_{oo^{\prime}}=\left\{
\begin{array}
[c]{l}%
-\frac{1}{3}\left(  1-A_{o}^{2}-2\sum_{o^{\prime\prime}=1}^{o-2}%
A_{o^{\prime\prime}}^{2}+2\sum_{e^{\prime\prime}=2}^{o-1}A_{e^{\prime\prime}%
}^{2}-2A_{o}B_{o}\right)  ,~for~o=o^{\prime}~\\
-\frac{1}{3}\sqrt{oo^{\prime}}\left(  \frac{\left(  A\bar{B}+B\bar{A}\right)
_{oo^{\prime}}}{o+o^{\prime}}-\frac{\left(  A\bar{B}-B\bar{A}\right)
_{oo^{\prime}}}{o-o^{\prime}}\right)  ,\;for~o\neq o^{\prime}%
\end{array}
\right. \\
\tilde{M}_{ee^{\prime}}  &  =\left(  \frac{1-\bar{t}t}{1+3t\bar{t}}\right)
_{ee^{\prime}}=\left\{
\begin{array}
[c]{l}%
-\frac{1}{3}\left(  1+A_{e}^{2}+2\sum_{e^{\prime\prime}=2}^{e-2}%
A_{e^{\prime\prime}}^{2}-2\sum_{o^{\prime\prime}=1}^{e-1}A_{o^{\prime\prime}%
}^{2}-2A_{e}B_{e}\right)  ,~for~e=e^{\prime}~\\
+\frac{1}{3}\sqrt{ee^{\prime}}\left(  \frac{\left(  A\bar{B}+B\bar{A}\right)
_{ee^{\prime}}}{e+e^{\prime}}-\frac{\left(  A\bar{B}-B\bar{A}\right)
_{ee^{\prime}}}{e-e^{\prime}}\right)  ,\;for~e\neq e^{\prime}%
\end{array}
\right. \\
\tilde{Y}_{eo}  &  =\left(  t\frac{1}{1+3t\bar{t}}\right)  _{eo}=\frac
{\sqrt{eo}}{4\sqrt{3}}\left(  \frac{\left(  A\bar{B}-B\bar{A}\right)  _{eo}%
}{e+o}-\frac{\left(  A\bar{B}+B\bar{A}\right)  _{eo}}{e-o}\right)  .
\label{mtilde3}%
\end{align}
The generating functions give the following values for $A_{e},A_{o}%
,B_{e},B_{o}$ which are useful in the present paper
\begin{align}
\sum A_{e}z^{e}  &  =\allowbreak-\frac{2}{3^{2}}z^{2}+\frac{2\times19}{3^{5}%
}z^{4}-\frac{2\times409}{3^{8}}z^{6}+\frac{2\times11\times283}{3^{10}}%
z^{8}-\frac{2\times220\,721}{3^{14}}z^{10}+\cdots\\
\sum A_{o}z^{o}  &  =\frac{2}{3}z-\frac{2\times11}{3^{4}}z^{3}+\frac
{2\times67}{3^{6}}z^{5}-\frac{2\times1409}{3^{9}}z^{7}+\frac{2\times
94\,993}{3^{13}}z^{9}+\cdots\\
\sum B_{e}z^{e}  &  =-\frac{2^{3}}{3^{2}}z^{2}+\frac{2^{4}11}{3^{5}}%
z^{4}-\frac{2^{3}523}{3^{8}}z^{6}+\frac{2^{5}29\times37}{3^{10}}z^{8}%
-\frac{2^{3}323\,381}{3^{14}}z^{10}+\cdots\\
\sum B_{o}z^{o}  &  =\frac{2^{2}}{3}z-\frac{2^{2}17}{3^{4}}z^{3}+\frac
{2^{2}127}{3^{6}}z^{5}-\frac{2^{2}11\times277}{3^{9}}z^{7}+\frac{2^{2}%
23\times9839}{3^{13}}z^{9}+\cdots
\end{align}
By applying these formulas we verified that the expressions in terms of the
vectors $A_{n},B_{n}$ given in Eqs.(\ref{bb}-\ref{mtilde3}) agree with the
results produced by the integrals in Eqs.(\ref{intf}-\ref{intvv}) as listed in
Eqs.(\ref{be}-\ref{Mtildeo}).

\end{document}